\documentclass[12pt]{article}

\usepackage{amsmath,amssymb,epsfig}
\usepackage[active]{srcltx}
\usepackage[all]{xypic}

\newcommand{\be}{\begin{equation}}
\newcommand{\ba}{\begin{eqnarray}}
\newcommand{\ea}{\end{eqnarray}}
\newcommand{\nn}{\nonumber}
\newcommand{\eproof}{{~\hfill$ \triangleleft$}}

\def\G{\Gamma}

\def\ca{{\cal A}}
\def\cb{{\cal B}}

\def\ce{{\cal E}}

\def\cg{{\cal G}}
\def\ch{{\cal H}}

\def\ck{{\cal K}}

\def\cn{{\cal N}}

\def\cp{{\cal P}}

\def\cs{{\cal S}}
\def\ct{{\cal T}}

\newtheorem{thm}{Theorem}[subsection]

\newtheorem{cor}[thm]{Corollary}
\newtheorem{prop}[thm]{Proposition}
\newtheorem{lemma}[thm]{Lemma}

\newtheorem{definition}[thm]{Definition}
\newtheorem{proposition}[thm]{Proposition}

\newcommand{\bbC}{{\Bbb C}}

\newcommand{\bbR}{{\Bbb R}}

\newcommand{\bbZ}{{\Bbb Z}}

\newcommand{\cG}{{\cal G}}

\fontfamily{yfrak}

\begin{document}
\vskip 15mm

\begin{center}

{\Large\bfseries On Spectral Triples in
Quantum Gravity II\\[2mm]  
}

\vskip 4ex

Johannes \textsc{Aastrup}$\,^{a}$\footnote{email: \texttt{johannes.aastrup@uni-muenster.de}},
Jesper M\o ller \textsc{Grimstrup}\,$^{b}$\footnote{email: \texttt{grimstrup@nbi.dk}}\\ \& Ryszard \textsc{Nest}\,$^{c}$\footnote{email: \texttt{rnest@math.ku.dk}}

\vskip 3ex  

$^{a}\,$\textit{SFB 478 ``Geometrische Strukturen in der Mathematik''\\
           Hittorfstr. 27, 48149 M\"unster, Germany}
\\[3ex]
$^{b}\,$\textit{The Niels Bohr Institute \\Blegdamsvej 17, DK-2100 Copenhagen, Denmark}
\\[3ex]
$^{c}$ \textit{Matematisk Institut\\ Universitetsparken 5, DK-2100 Copenhagen, Denmark}
\end{center}

\vskip 5ex

\begin{abstract}

A semifinite spectral triple for an algebra canonically associated to canonical quantum gravity is constructed. The algebra is generated by based loops in a triangulation and its barycentric subdivisions. The underlying space can be seen as a gauge fixing of the  unconstrained state space of Loop Quantum Gravity.   
This paper is the second of two papers on the subject.
\end{abstract}

\newpage

\tableofcontents

\section{Introduction}

One of the most striking application of Noncommutative Geometry \cite{Co} is Connes' derivation of the Standard Model of high energy physics \cite{Co2,CCM}. In this derivation the Lagrangian of the full Standard Model coupled to gravity emerges from the spectral action principle \cite{AC1} applied to a specific almost commutative geometry. 

This formulation is, however, essentially classical and does, at a fundamental level, not involve quantization. This raises the question how the quantization procedure of Quantum Field Theory fits into the framework on Noncommutative Geometry. Since Connes' formulation of the Standard Model at its root is tied up to gravitation, a quantization scheme within the framework of Noncommutative Geometry must be expected to involve, at some level, quantum gravity.

In the papers \cite{AG1} and \cite{AG2} we  started studying the question of formulating a quantization scheme within  Noncommutative Geometry using canonical quantum gravity. Specifically, we were inspired by techniques applied in Loop Quantum Gravity. The concrete aim was to construct a spectral triple of a noncommutative algebra naturally associated to the unconstrained state space of Loop Quantum Gravity. Due to technical difficulties we only succeeded in constructing a spectral triple on a space closely related to the state space of Loop Quantum Gravity.

This paper is  one of two papers presenting a more satisfactory solution to the problem of constructing a spectral triple on the state space of Loop Quantum Gravity as well as its physical interpretations. The paper \cite{AGN1} deals with the physical background and interpretations of the construction, and we therefore refer the reader to that paper for more thorough discussion. This paper deals with the concise mathematical construction.  

\subsection{Content of the paper}  
 The unconstrained state space of Loop Quantum Gravity is the space of $SU(2)$-connections in a trivial principal bundle on a three dimensional manifold. Since the construction we do works for  arbitrary manifolds  and arbitrary compact Lie-groups, and since we for physical applications might need it for later use, we will formulate it in this generality.

We let $G$ be a compact Lie group and assume we have some principal $G$-fiber bundle $P$ over a manifold $M$. The algebra associated to the space $\ca$ of connections in $P$  we want to consider is the following:

Given a representation of $G$ in $M_N$ and given a loop $L$ in $M$ define a matrix valued function on $\ca$ by
$$L(\nabla )=Hol(\nabla ,L),$$
where $Hol(\nabla ,L)$ denotes the holonomy of $\nabla$ along $L$. Let $\cb$ be the algebra generated by all loops based in a fixed point in $M$. This is the type of algebra we want to consider. 

The algebra $\cb$ is an algebra of matrix valued functions over $\ca$. We will for technical reasons explained below consider a  smaller algebra than $\cb$. We will take a triangulation of $M$ and consider the algebra $\cb_\triangle$ which is constructed like $\cb$ but only includes loops lying in the edges of the triangulations, or one of its barycentric subdivisions. $\cb_\triangle$ is also an algebra of matrix valued functions over $\ca$. Since any loop can be approximated by a loop lying in the edges of the triangulations, or one of its barycentric subdivisions, $\cb_\triangle$ can be considered as an approximation to $\cb$. The crucial difference is that the groups of all diffeomorphisms preserving the base point acts on $\cb$ whereas only the group differomorphisms preserving the chosen triangulation and its barycentric subdivisions acts on $\cb_\triangle$.


In order to construct a spectral triple over $\cb_\triangle$ we need a representation of $\cb_\triangle$. We therefore need to construct $L^2(\ca )$. 
A construction of  $L^2(\ca )$ already exists within Loop Quantum Gravity, due to Ashtekar and Lewandowski. It turns out that this construction depends on a completion of $\ca$ and that this completion depends on a choice of a system $\cs$ of graphs. In the case of $\cb_\triangle$ the relevant system of graphs is given by finite subgraphs of the triangulation and its barycentric subdivision. The system of graphs considered in Loop Quantum Gravity is that of graphs made up of piecewise  real analytic edges (of course assuming $M$ to be real analytic). 

Seen from a graph $\Gamma$ with edges $\{e_i\}_{i=1,\ldots ,n}$ the space of connections $\ca$ looks like $G^{n}$ via
$$\ca \ni \nabla \to (Hol(\nabla, e_1 ),\ldots ,Hol(\nabla ,e_{n})).$$
We will also denote $\ca_\Gamma :=G^n$. Of course $\ca_\Gamma$ tells little about $\ca$. However, by letting the complexity of the graph grow we get a more and more refined picture of $\ca$. This is implemented by noting that an embedding $\Gamma_1 \subset \Gamma_2$ naturally gives a map
$$P_{\Gamma_1\Gamma_2}:\ca_{\Gamma_2}\to \ca_{\Gamma_1},$$
and then simply define the completion of $\ca$ as the projective limit of $\ca_\Gamma$, i.e.
$$\overline{\ca}^\cs:=\lim_{\stackrel{\Gamma \in \cs}{\longleftarrow} }\ca_\Gamma,$$
We will prove that under some condition on $\cs$ that $\ca$ is densely embedded in  $\overline{\ca}^\cs$, and hence justifies the term completion.

The construction of $L^2(\ca)$ is straightforward from this completion. Since $\ca_\Gamma =G^n$ we define $L^2(\ca_\Gamma )$ as square integrable functions on $G^n$ with respect to the Haar measure. The map $P_{\Gamma_1 \Gamma_2}$ induces an embedding
$$P_{\Gamma_1 \Gamma_2}:L^2 (\ca_{\Gamma_1})\to L^2(\ca_{\Gamma_2}),$$
and $L^2(\ca)$ is defined as an inductive limit
$$L^2(\ca ):=\lim_{\stackrel{\Gamma \in \cs}{\longrightarrow} }L^2(\ca_\Gamma).$$


The idea for constructing the Dirac operator is  that $\ca_\Gamma =G^n$ is a classical geometry and therefore has a canonical Gau\ss-Bonnet-Dirac operator. In order to ensure that this defines an operator on the inductive limit we have to make sure that the operator is compatible with the structure maps of the projective limit. It is here the technical advantages of the triangulation compared to piecewise real analytic graphs shows up. The triangulation narrows down the types of structure maps appearing in the projective limit. With this we can  define a Dirac operator compatible with the structure maps. It turns out that there is a lot freedom in the construction. Going from one level in the inductive limit to the next level we add new copies of $G$ which corresponds to new degrees of freedom. Each of these degrees of freedom can be scaled. The entire construction therefore comes with a sequence of real non zero numbers $\{ a^{j,k}\}_{k\leq j}$, where $j,k$ is just a labelling of the degrees of freedoms convenient for the explicit construction. 

However, the constructed Dirac operator together with the algebra $\cb_\triangle$  will not fulfil the requirement of a spectral triple. The problem comes from the infinite dimensionality of the Clifford bundle of the space of connection. We therefore need to treat the identity operator on the Clifford bundle as a finite rank operator. The setting of  semifinite spectral triples will allow exactly this. The von Neumann algebra $\cn$ appearing from this construction  is a tensor product of the bounded operators on a separable Hilbert space and the weak closure of the CAR-algebra. 

In order to obtain a semifinite spectral triple we still need to perturb the operator. At each level in the inductive limit we have to add a bounded perturbation roughly speaking of the form  $b_jP_j$, where $b_j$ is a real number and $P_j$ is the projection onto a subspace related to the kernel of a part of the Dirac operator at the $j$'th level. 

The main result in the paper is \ref{hoved} stating\\
\textbf{Theorem}\textit{
There exist sequences $\{ a^{j,k} \}$ and $\{ b_j\}$ such that $\cb_\triangle$ together with the perturbated Dirac operator  is a semifinite spectral triple with respect to the trace on $\cn$.}\\

In the appendix we will demonstrate the case of $U(1)$ and show that for $SU(2)$ the perturbations are not needed.\\

\noindent{\bf Acknowledgements}\\
We thank the following colleagues for fruitful discussions: Christian Fleischhack, Victor Gayral, Gerd Grubb, Troels Harmark, Thordur Jonsson, Mario Paschke, Adam Rennie, Carlo Rovelli, Thomas Schucker, Christian Voigh and Raimar Wulkenhaar. Furthermore, we are grateful to the following institutes for hospitality during visits: the Institute of Mathematics in Reykjavik, Iceland; The Max Planck Institute for Mathematics in the Sciences, Leipzig, Germany; the Isaac Newton Institute for Mathematical Sciences, Cambridge, UK; the Institute of Theoretical Physics in Marseilles, France.

Johannes Aastrup was  funded by the German Research Foundation (DFG) within the research projects  {\it Deformation Theory for Boundary Value Problems} and {\it Geometrische Strukturen in der Mathematik} (SFB 478).

\section{Spaces of connections} \label{sam}
 In this section we will consider the space $\ca$ of connections in a principal fiber bundle. We will address the problem of topologizing this space together with the development of a measure theory. It turns out that the constructions, which we introduce in order to address these problems, depends on a completion of $\ca$ with respect to a system of graphs on the manifold. Different choices of graphs give different completions.  

The material is standard from Loop Quantum Gravity, but we have chosen to write a more or less selfcontained exposition, not assuming prior knowledge of Loop Quantum Gravity, since we want to put emphasize on several different completions and their mutual interplay.  The original techniques were developed by Ashtekar and Lewandowski in \cite{AL2}. For a survey of Loop Quantum Gravity see \cite{AL1} and for a detailed account see \cite{Th1}.\\

Let $M$ be a manifold and $P$ a $G$-principal bundle over $M$, where $G$ is a compact connected Lie-group. We will for simplicity assume that $P$ is the trivial bundle, i.e. $P$ is isomorphic to $M\times G$.

\subsection{Graphs}
\begin{definition}\label{curve}
Let $\gamma$ be  a  continuous piecewise smooth map $\gamma:[0,1]\to M$ such that if $\gamma (t_1)=\gamma(t_2)$ then $t_1=t_2$ or $t_1,t_2\in \{ 0,1 \}$. We call $\gamma$ a simple curve if $\dot{\gamma}(t)$ is non vanishing for all $t\in [ 0,1]$. This requirement includes that the left and right derivative of $\gamma$ is non vanishing in the non smooth points. 
\end{definition}

We will call $\gamma (0)$ the starting point and $\gamma (1)$ the endpoint.

If $\gamma_1$ and $\gamma_2$ are simple curves where the endpoint of $\gamma_1$ coincides with the starting point of $\gamma_2$, then the composition is defined by
$$(\gamma_1\circ \gamma_2) (t)=\left\{ 
\begin{array}{cl} 
\gamma_1(2t)& t\in [0,\frac{1}{2}]\\
\gamma_2(2t-1) & t\in [\frac{1}{2},1]
\end{array}\;.
\right.$$
 Note that the composition of two simple curves is not always a simple curve.

\begin{definition}\label{path}
Two simple curves $\gamma_1$ and $\gamma_2$ are equivalent if there exist an increasing piecewise smooth bijection
$$\phi :[0,1]\to [0,1]$$
with 
$$\gamma_1 (\phi (t))=\gamma_2(t), \quad t \in [0,1],$$
and with nowhere vanishing derivative. This requirement includes that the left and right derivatives in the non smooth points are both non vanishing. 

An equivalence class of simple curves is called a simple path.
\end{definition}

Note that the relation defined above is in fact an equivalence relation.  

The notion of a simple path  chosen here implies that a simple path has an orientation. The inverse of a simple path represented by $\gamma$ is the path represented by $\gamma^{-1}$, where
$$\gamma^{-1}(t)=\gamma (1-t),$$
i.e. just the path with inverse orientation.

\begin{definition}\label{graph}
A graph $\Gamma$ is the union of finite many simple paths  $\{ e_i\}$ where the only intersection points are start or end points of the simple paths. 

The paths are called the edges of the graph. We denote the set of edges of $\Gamma$ by $E_\Gamma$. The points $\{ e_i(0) ,e_j(1)\}$ are called the vertices of $\Gamma$. We denote the set of vertices by $V_\Gamma$.

A graph is called connected if $\cup_{i,t}e_i (t)$ is connected.
\end{definition}

We will consider paths on the graph. These are simply compositions of edges and inverses of edges. We will also think of each vertex as a path in the graph. Furthermore, we would like $e_i \circ (e_i)^{-1}$ to be equal to the  path $e_i(0)$. We fix this with the following 
\begin{definition}\label{pathgroupoid}
Let $\Gamma$ be a graph and let $\cp (\Gamma )$ be the set of paths in $\Gamma$. We define an equivalence relation $\sim$ on $\cp (\Gamma )$ to be generated by 
$$(e_i)^{-1}\circ e_i \sim e_i(1)\hbox{ and } e_i\circ (e_i)^{-1} \sim e_i(0).$$

The hoop groupoid  $\ch\cG(\Gamma)$  of path on $\Gamma$ is defined by 
$$\ch\cg(\Gamma ) = \cp (\Gamma )/ \sim.$$

\end{definition}

This equivalence relation implies that if $p_1,p_2,p_3\in \cp(\Gamma )$ and $p_1 \sim p_2$ then $p_3\circ p_1\sim p_3 \circ p_2$ and $p_1\circ p_3\sim p_2\circ p_3$ whenever the compositions are defined. The units for $\ch\cg (\Gamma )$ are the vertices of $\Gamma$. It is easy to see that $\ch \cg (\Gamma )$ is a groupoid.

\begin{definition}\label{subgraph}
Let $\Gamma_1 ,\Gamma_2$ be two graphs. We say that $\Gamma_1$ is a subgraph of $\Gamma_2$, and we write $\Gamma_1\subset \Gamma_2$, if the edges of $\Gamma_1$ are compositions of the edges or their inverses of $\Gamma_2$.
\end{definition}
Note that $\Gamma_1 \subset \Gamma_2$ implies that $\ch \cg(\Gamma_1)$ is a subgruopoid of $\ch\cg (\Gamma_2 )$.

The relation $\subset$ equips the set of graphs with a partial order. We are interested in subsystems of the set of all graphs.

\begin{definition}
A system $\cs$ of graphs is called directed if there for any $\Gamma_1,\Gamma_2 \in \cs$ exists $\Gamma_3 \in \cs$ with $\Gamma_1 ,\Gamma_2 \subset \Gamma_3$.
\end{definition}
\begin{definition}
A system of graphs $\cs$ is called dense if there for every point $m\in M$ exists a coordinate chart ${\bf{x}}=(x_1,\ldots ,x_n)$ around $m$ such that for all open subset $U$ containing $m$ in this coordinate chart there exists a collection of edges $e_1,\ldots , e_n\subset U$ belonging to graphs in $\cs$ such that:
\begin{enumerate}
\item 
 the $e_i$'s are straight lines with respect to the coordinate chart,
\item
the tangent vectors of the $e_i$'s are linearly independent.
\end{enumerate}
\end{definition}

The definition of dense might appear awkward. The main purpose of the definition is to ensure proposition \ref{dense}. Furthermore it is easy to check the definition in the concrete examples we have in mind. Certainly the requirement of straight lines can be eased.

 We will now give three examples of dense systems of graphs, which will be important in the rest of the paper. These systems of graphs differ particularly in the size of their corresponding symmetry groups.\\

\textbf{Example 1} Let $\cs_s$ be the system of all graphs. This system is clearly dense. The system is however not directed since we can have two simple paths $e_1, e_2$ with infinitely many isolated intersection points. Hence for graphs $\Gamma_1 ,\Gamma_2$, where $e_1\in \Gamma_1$ and $e_2\in \Gamma_2$, there does not exist a graphs $\Gamma_3$ containing $\Gamma_1$ and $\Gamma_2$. The system $\cs_s$ admits a natural action of the diffeomorphism group $Diff(M)$. The system $\cs_s$ is the same as the piecewise immersed system defined in \cite{Fl1}.\\

\textbf{Example 2} Let $M$ be a real analytic manifold. Let $\cs_a$ be the system of graph made up of real analytic simple curves. This system is dense and it is also directed since piecewise analytic curves have only finitely many isolated intersection points. The system carries a natural action of the group $Diff_a(M)$ of real analytic diffeomorphism, but no action of the full diffeomorphism group $Diff (M)$. This system was, in a base pointed version, first considered in \cite{AL2}.\\

\textbf{Example 3} Let  $\ct$ be a triangulation of $M$. We let $\Gamma_0$ be the graph consisting of all the edges in this triangulation. Strictly speaking this is not a graph if the manifold in not compact, but in this case we can consider $\Gamma_0$ as a system of graphs instead. Let $\ct_n$ be the triangulation obtained by barycentric subdividing each of the simplices in $\ct$ $n$ times. 
 The graph $\Gamma_n$ is  the graph consisting of the edges of $\ct_n$. In this way we get a directed and dense system  $\cs_\triangle =\{ \Gamma_n\}$ of graphs. \\ 

The important feature of $\cs_\triangle$ which we are going to use in this paper is the following: 
{\textbf{The step from $\Gamma_n$ to $\Gamma_{n+1}$ involves:}
\begin{enumerate}
\item  
{\textbf{new edges are added}}
\item 
{\textbf{the edges of $\Gamma_n$ get subdivided into two edges.}}
\end{enumerate}

The system $\cs_\triangle$ only admits an action of the diffeomorphisms $Diff (\triangle )$ that maps edges in $\cup \ct_n$ to edges in $\cup \ct_n$. Hence this is a much more restrictive class of diffeomorphism than in the first two examples. Contrary to the first two examples, the system $\cs_\triangle$ is countable.

\begin{figure} [t]
\begin{center}
 \input{JohannesFig.pstex_t}
\caption{}
\label{JohannesFig}
\end{center}
\end{figure}

\begin{definition}
Let $\cs$ be a directed system of graphs. We define the hoop groupoid $\ch\cg (\cs)$ of $\cs$ to be the inductive limit 
$$\ch\cg (\cs ) =\lim_{\stackrel{\Gamma \in \cs}{\longrightarrow}} \ch\cG (\Gamma ).$$  
\end{definition}

\subsection{Completing spaces of connections} 
 Given a graph $\Gamma$ define the space 
\[
\ca_\Gamma = Hom(\ch\cg(\Gamma),G)
\]
where $G$ is the compact connected Lie-group.
If $\Gamma_1 \subset \Gamma_2$ we have the embedding of groupoids $\ch\cg(\Gamma_1)\to \ch \cg (\Gamma_2)$, and we hence get a surjection
$$P_{\Gamma_1\Gamma_2}: \ca_{\Gamma_2} \to \ca_{\Gamma_1}. $$
Therefore, for a system of graphs $\cs$ we have a projective system 
$$\{ \ca_\Gamma \}_{\Gamma \in \cs}.$$
 
\begin{definition}
Let $\cs$ be a system of graphs. The space of generalized connections with respect to $\cs$, denoted by $\overline{\ca}^\cs$, is defined by
$$\overline{\ca}^\cs= \lim_{\stackrel{\Gamma \in \cs}{\longleftarrow} }\ca_\Gamma.$$ 
The projections from $\overline{\ca}^\cs$ to $\ca_\Gamma$ will be denoted by $P_\Gamma$.

\end{definition}
For the systems $\cs_s, \cs_a, \cs_\triangle$ we will denote the corresponding spaces of generalized connections by $\overline{\ca}^s,\overline{\ca}^a,\overline{\ca}^\triangle$.

Note that when $\cs$ is directed we have the equality $$\overline{\ca}^\cs=Hom(\ch\cg (\cs),G).$$

Given an element $\nabla$ in $\ca_\Gamma$ we can associate to it $\Phi_\Gamma (\nabla )\in G^{ n_\Gamma}$ where $n_\Gamma$ is the numbers of edges in $\Gamma$. This is done by numbering the edges in $\Gamma$ as $e_1,\ldots , e_{n_\G}$, and then defining $\Phi_\Gamma$ by
$$\Phi_\Gamma (\nabla )=(\nabla (e_1),\ldots , \nabla (e_{n_\Gamma})).$$

\begin{lemma} \label{gn}
The map $\Phi_\Gamma : \ca_\Gamma \to G^{n_\Gamma}$ is a bijection.
\end{lemma}

\textit{Proof.} Follows since $\ch \cg (\Gamma)$ is freely generated by the edges. \eproof\\

The bijection $\Phi_\Gamma$ gives a topology on $\ca_\Gamma$ by requiring $\Phi_\Gamma$ to be a homeomorphism. The topology is independent of the chosen numbering.  The projection maps $P_{\Gamma_1 \Gamma_2}:\ca_{\Gamma_2}\to \ca_{\Gamma_1}$ are continuous. In fact 

$$\Phi_{\Gamma_1}\circ P_{\Gamma_1 \Gamma_2} \circ (\Phi_{\Gamma_2})^{-1} : G^{n_{\Gamma_2}}\to G^{n_{\Gamma_1}}$$
is given by composition of one or more of the following operations:
\begin{itemize}
\item Multiplying $g_{i_1}$ and $g_{i_2}$.
\item Inverting $g_i$.
\item Leaving out some $g_i$ in $(g_1,\ldots , g_{n_{\Gamma_2}})\in G^{n_{\Gamma_2}}$.
\end{itemize}
 
Since $\overline{\ca}^\cs$ is a projective limit of $\{ \ca_\Gamma\}_{\Gamma \in \cs}$, we define the topology on $\overline{\ca}^\cs$ as the projective limit topology. This topology is characterized by the following property:

Let $X$ be a topological space and assume we have continuous maps $\phi_\Gamma :X\to \ca_\Gamma$ for all $\Gamma \in \cs$ such that $P_{\Gamma_1\Gamma_2}\circ \phi_{\Gamma_2}= \phi_{\Gamma_1}$ for all $\Gamma_1\subset \Gamma_2$. Then there is a unique continuous map $\phi: X\to \overline{\ca}^\cs$ with $P_\Gamma \circ \phi_\Gamma=\phi$ for all $\Gamma\in\cs$.

\subsubsection{Smooth connections} Let $\ca$ denote the space of smooth connections in the principal bundle $P$. There is a natural map 
$$\chi_\Gamma:\ca \to \ca_\Gamma=Hom (\ch\cg(\Gamma ),G)$$ defined by
$$\chi_\Gamma (\nabla )(p )=Hol (p,\nabla ), \quad p\in \ch\cg(\Gamma ),$$
where $Hol(p,\nabla )$ denotes the holonomy of $\nabla$ along $p$. Clearly $P_{\Gamma_1\Gamma_2} \circ\chi_{\Gamma_2}=\chi_{\Gamma_1}$ when $\Gamma_1\subset \Gamma_2$, and hence by the property of the projective limit we get a unique map
$$\chi_\cs :\ca\to \overline{\ca}^\cs.$$

\begin{prop}
When $\cs$ is directed, $\chi_\cs (\ca)$ is dense in $\overline{\ca}^\cs$.
\end{prop}

\textit{Proof.} We first prove that each $\chi_\Gamma$ is surjective. The composition 
$$\Phi_\Gamma \circ\chi_\Gamma:\ca \to G^{n_\Gamma}$$
is given by 
$$\Phi_\Gamma (\chi_\Gamma (\nabla))=(Hol(e_1,\nabla),\ldots , Hol (e_{n_\Gamma},\nabla)).$$
Let $(g_1,\ldots, g_{n_\Gamma})$ be given. Since $G$ is connected we can for each $i$ find a connection $\nabla_i$ with 
$$Hol(e_i , \nabla_i)=g_i.$$
It is furthermore easy to see that since the $e_i$'s only intersect in the endpoints we can arrange that $\nabla_i=0$ on all the edges in $\Gamma$  apart from $e_i$. Hence 
$$\Phi_\Gamma\left(\chi_\Gamma \left(\sum \nabla_i \right) \right)=(g_1,\ldots , g_{n_\Gamma }).$$
Therefore $\chi_\Gamma$ is surjective.
 
From the directedness of $\cs$ and the surjectivity of all $\chi_\Gamma$ the density follows. \eproof\\

The system $\cs_s$ is not directed. It is however possible with more elaborate methods to prove that $\ca$ is dense in $\overline{\ca}^{\cs_s}$ when $G$ is semi-simple, see for example the discussion in \cite{Fl2}. 

We now turn to the question of injectivity of  $\chi$.
\begin{prop}\label{dense}
When $\cs$ is dense, $\chi_\cs$ is injective.
\end{prop}

\textit{Proof.} Let $\nabla_1,\nabla_2$ be two different smooth connections. Hence there is a point $m\in M$ where $\nabla_1 (m) \ne \nabla_2 (m)$. Let us choose a coordinate chart ${\bf{x}}=(x_1,\ldots , x_n)$ around $m$ according to the density of $\cs$, where $m$ corresponds to $\bf{x}=0$. We can then write
$$\nabla_k=\sum_j \mathfrak{g}_j^k({\bf{x}})dx_j, \quad k\in \{ 1,2\},$$
where $ \mathfrak{g}_j^i({\bf{x}})$ is a smooth function of $\bf{x}$ with values in the Lie-algebra of $G$. Let $U$ be a neighbourhood of $\bf{0}$ such that we can assume that the functions $ \mathfrak{g}_j^i({\bf{x}})$ are constant in $U$ with sufficiently good approximation. Let $e_1,\ldots , e_n$ be the edges in $\cs$ which are straight lines with respect to the coordinate chart, which belongs to $U$ and where the tangent vectors ${\bf{t}}^1,\ldots,{\bf{t}}^n$ are linearly independent. We assume that the edges are parametrized by arc lengths.   Since $\nabla_1(m)\ne \nabla_2(m)$ there is a $j$ such that 
$$ \sum_i \mathfrak{g}_i^1({\bf{0}})t^j_i \ne \sum_i\mathfrak{g}_i^2({\bf{0}})t^j_i.  $$ 
 With sufficiently good approximation we have 
$$Hol(e_j ,\nabla_k)=\exp ( \sum_i \mathfrak{g}_i^1({\bf{0}})t^j_i),\quad k\in \{1,2\}.$$
Hence, if we had chosen $U$ small enough, we conclude 
 $$Hol(e_j,\nabla_1)\ne Hol (e_j,\nabla_2).$$
Thus $\chi_\cs (\nabla_1)\ne \chi_\cs (\nabla_2)$.\eproof

\subsubsection{Completing the group of gauge transformations} \label{gauge}
Let $U$ be a an element in the gauge group $\cg$ of $M\times P$, i.e. $U:M\to G$ is a smooth function. Given a connection $\nabla \in \ca$, $U$ induces a gauge transformed connection $\tilde{\nabla}$. Given a path $p$ on $M$ with startpoint $x_0$ and endpoint $x_1$ the holonomy along $e$ transforms according to 
$$Hol (p,\tilde{\nabla})=U(x_0)Hol(p,\nabla )U^{-1}(x_1).$$
This leads to the following 
\begin{definition}
 Let $\Gamma$ be a graphs and $U:V_\Gamma \to G$  be a map. Define $U_*:\ca_\Gamma \to \ca_\Gamma$ by
$$U_*(\nabla )(e)=U(e(0))\nabla (e)U(e(1))^{-1}\hbox{ for all } e\in E_\Gamma.$$
Since $\ch\cg (\Gamma)$ is freely generated by $E_\Gamma$, this is well defined.

We denote by $\cg_\Gamma$ the group of all maps $U:V_\Gamma \to G$.
\end{definition}

Note that via $U_*$ we get a left action of $\cg_\Gamma$ on $\ca_\Gamma$. 

Like with spaces of connections there are natural projections $$P_{\Gamma_1 \Gamma_2}:\cg_{\Gamma_2} \to \cg_{\Gamma_1},$$ 
when $\Gamma_1 \subset \Gamma_2$ and 
\begin{equation} \label{kop}
P_{\Gamma_1 \Gamma_2}(U_*(\nabla ))=P_{\Gamma_1 \Gamma_2}(U)_*(P_{\Gamma_1\Gamma_2}(\nabla)),\quad \nabla\in \ca_{\Gamma_2},U\in \cg_{\Gamma_2}.
 \end{equation}

\begin{definition}
Let $\cs$ be a system of graphs. Put 
$$\overline{\cg}^\cs=\{ U |U:\cup_{\Gamma \in \cs} V_\Gamma \to G\}.$$
Define the left action of $\overline{\cg}^\cs$ on $\overline{\ca}^\cs$ by
$$U \cdot \nabla= (U|_{V_\Gamma})_*(\nabla ),\quad \nabla\in \ca_\Gamma .$$  
\end{definition}

Due to (\ref{kop}) this is well defined.

A gauge transformation $U\in \cg$ naturally gives an element in $\overline{\cg}^\cs$. When $\cs$ is dense this induces an embedding $\cg \to \overline{\cg}^\cs$. Via this embedding we get an action of $\cg$ on $\overline{\ca}^\cs$, which extends the action of $\cg$ on $\ca$. We will therefore call $\overline{\cg}^\cs$ the completed gauge group, or simply the gauge group.

\subsection{Group actions on the completions}
If a system $\cs_1$ is contained in $\cs_2$ we get a surjective continuous map 
$$P_{\cs_1 \cs_2}:\overline{\ca}^{\cs_2} \to \overline{\ca}^{\cs_1}.$$
In particular we have the commutative diagram
$$\hbox{\xymatrix{\overline{\ca}^s \ar@/_1pc/[rr]_{P_{\cs_\triangle \cs_s}}  \ar[r]^{P_{\cs_a \cs_s}} &\overline{\ca}^a  \ar[r]^{P_{\cs_\triangle \cs_a}} &\overline{\ca}^\triangle}}$$
Of course the map $P_{\cs_\triangle \cs_a}$ only exists if the triangulation is real analytic.

The group of all diffeomorphisms $Diff (M)$ acts on $\ca$. The question is  what kind of groups  acts on the different completions $\overline{\ca}^\cs$. In general a diffeomorphism $d$ preserving $\cs$ induces a homeomorphism $d^*:\overline{\ca}^\cs \to \overline{\ca}^\cs$. Thus in this way there is an action of $Diff(\cs )$, the diffeomorphisms preserving $\cs$, on  $\overline{\ca}^\cs$. We have the following diagram
$$
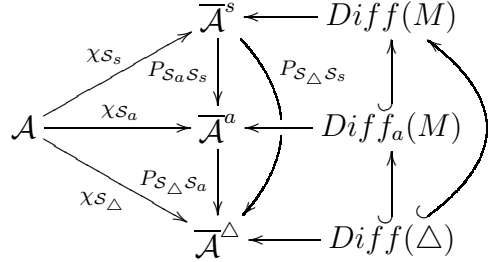
\hbox{\xymatrix{ & &\overline{\ca}^s \ar[d]_{P_{\cs_a \cs_s}} \ar@/^2pc/[dd]^(0.3){P_{\cs_\triangle \cs_s}} |!{[d];[dr]}\hole& \ar[l] Diff(M)\\
\ca \ar[urr]^{\chi_{\cs_s}} \ar[drr]_{\chi_{\cs_\triangle}} \ar[rr]^{\chi_{\cs_a}}&& \overline{\ca}^a \ar[d]_{P_{\cs_\triangle \cs_a}}& Diff_a(M) \ar[l] \ar@{^{(}->}[u]  \\
&& \overline{\ca}^\triangle & Diff (\triangle )\ar[l]  \ar@{^{(}->}[u] \ar@{^{(}->}@/_3pc/[uu] }}     $$
We see that the size of the completion of $\ca$ is strongly related to the size of the symmetry group. The spaces $\overline{\ca}^s,\overline{\ca}^a$ are non separable and the symmetry groups are large, whereas $\overline{\ca}^\triangle$ is separable and the symmetry group $Diff (\triangle )$ is comparatively small. In this way we can think of $\overline{\ca}^\triangle$ as being $\overline{\ca}^s$ (or $\overline{\ca}^a$) subjected to a kind of gauge fixing of $Diff(M)$ (or $Diff_a (M)$).

\subsection{Measure theory and Hilbert spaces}
We will here recall the construction of measures and Hilbert space structures on completions of spaces of connections. The construction first appeared in  \cite{AL2}. See also \cite{MM} for a different approach.

Because of lemma \ref{gn} we can identify $\ca_\Gamma$ with $G^{n_\Gamma}$ via $\Phi_\Gamma$. On $G^{n_\Gamma}$ there is a canonical normalized measure, namely the Haar measure $\mu_{G^{n_\Gamma}}$. Denote by $\mu_\Gamma$ the image measure of $\mu_{G^{n_\Gamma}}$ under $\Phi_\Gamma^{-1}$. 

\begin{lemma} \label{maal}
Let $\Gamma_1 \subset \Gamma_2$. The image measure of $\mu_{\Gamma_2}$ under $P_{\Gamma_2\Gamma_1}$ is $\mu_{\Gamma_1}$. 
\end{lemma}  

\textit{Proof.} See for example lemma I.2.9 in \cite{Th1}. \eproof \\

Lemma \ref{maal} ensures that $\{ \ca_\Gamma ,\mu_\Gamma \}_{\gamma \in \cs}$ is a projective system of measure spaces. Therefore, according to Theorem I.2.10 in \cite{Th1}, there is a unique measure $\mu$ on $\overline{\ca}^\cs$ such that $P_\Gamma (\mu )=\mu_\Gamma$ for all $\Gamma \in \cs$. Also lemma \ref{maal} ensures that 
$$P^*_{\Gamma_2 \Gamma_1} :L^2(\ca_{\Gamma_1})\to L^2(\ca_{\Gamma_2})$$
is an  embedding of Hilbert spaces when $\Gamma_1 \subset \Gamma_2$. 
\begin{prop}\label{induktiv}
Let 
$$L^2(\ca)=\lim_{\longrightarrow}L^2(\ca_\Gamma ),$$
the inductive limit of the Hilbert spaces $\{ L^2 (\ca_\Gamma )\}_{\Gamma \in \cs}$. Then $L^2(\ca )=L^2(\overline{\ca}^\cs)$, where the latter is with respect to the measure $\mu$.
\end{prop}

\textit{Proof.} This follows directly from the construction. See for example section I.2.4 in \cite{Th1}. \eproof

\section{The Dirac operator on the completed space of connections} \label{dir}
In this section we will only work with the completed space $\overline{\ca}^\triangle$. We will construct a Dirac type operator acting on a Hilbert space $\ch$ which can naturally be seen as square integrable functions on $\overline{\ca}^\triangle$ with values in an infinite dimensional vector bundle.

The idea is to construct a Dirac operator on each $\ca_\Gamma$. Since these look like  classical geometries we more or less only have to put a Riemannian metric on each of these. Next we need to check that the construction made on each $\ca_\Gamma$ is consistent with maps between the different graphs. 

In this section we have no restrictions on the manifold $M$. In particular we can have infinitely many simplices in $\ct_0$.

To simplify the notation objects indexed by the  graph $\Gamma_k$ will be indexed simply by $k$ for the rest of the paper.

\subsection{The case of subdividing an edge into  two edges}
The construction of the Dirac operator, which will be carried out in the following subsections, looks cumbersome, but is really  forced upon us by the requirement of consistency with the maps between different graphs.  Therefore, in order to present the construction without the full notational weight, we will first demonstrate the case where the triangulation consists of one edge and where we consider the step going to the first barycentric subdivision. We thus have $\ca_0=G$, $\ca_1=G^2$ and $P_{0,1}(g_1,g_2)=g_1g_2$. 

For $\Gamma_0$ we choose a left and right invariant metric $\langle \cdot ,\cdot \rangle$ on $G$. Let $\hat{e}_i$ be an orthonormal basis for $T_{id}G$ and let $\hat{E}_i(g)=L_g\hat{e}_i$ be the corresponding left translated vector fields. Define the Dirac operator 
$$D_0(\xi) =\sum \hat{E}_i\nabla_{\hat{E}_i} (\xi),$$
where $\nabla$ is some $SO(\hbox{dim}(G))$-valued connection and $\xi \in L^2(G,Cl(TG))$, where $Cl(TG)$ denotes the Clifford bundle.

We want to construct $D_1$ acting on $L^2(G^2,Cl(G^2))$ on the form 
$$D_1(\xi) =\sum_{j} K_j \nabla_{K_j} (\xi), $$
where $K_j$ is an orthonormal frame in $TG^2$, such that $$P^*_{0,1} (D_0(\xi))=D_1(P_{0,1}^* (\xi)).$$ 

First we have to make sense of $P^*_{0,1}$. 
Since $P_{0,1}$ induces maps 
\[
(P_{0,1})_*:TG^2\rightarrow TG\;,\quad \mbox{and} \quad P_{0,1}^*:T^*G\rightarrow T^*G^2
\]
between tangent and cotangent spaces it is also natural to let the Dirac operators act on $Cl(T^*G)$, resp. $Cl(T^*G^2)$ instead of $Cl(TG)$, resp. $Cl(TG^2)$. This is easily done once we have chosen a metric on $T^*G$ and resp. $T^*G^2$. In order for 
$$P_{0,1}^* :L^2(G, Cl(T^*G)) \to L^2(G^2 ,Cl(T^*G^2))$$
to be an embedding of Hilbert spaces, and in fact to be defined at the level of Clifford bundles, the map induced by $P_{0,1}$ at the level of cotangent bundles, also denoted $P^*_{0,1}$, must be metric. 

Let $w \in T^*_gG$. It is easy to see that 
$$P^*_{0,1}(w )=(R_{g_2^{-1}} w , L_{g_1^{-1}}w ), \quad g_1g_2=g,$$
where 
$$R_{g_2^{-1}}(w)(v)=w (R_{g_2}v)$$ 
and 
$$L_{g_1^{-1}}(w )(v)=w(L_{g_1}v)$$. 
We ensure that $P_{0,1}^* :T^*G\to T^*G^2$ is metric by defining the inner product on $T^*G^2$ 
$$\langle (w_1,w_2),(w_3,w_4)\rangle_2 =\frac{1}{2}(\langle w_1,w_3\rangle +\langle w_2,w_4 \rangle ).$$

Denote by $E_i$ the cotangent vector field which is dual to $\hat{E}_i$ via $\langle \cdot , \cdot \rangle$. Using the inner product on $G^2$ we get from $P_{0,1}^*(E_i)$ a vector field on $G^2$. A small computation shows that this vector field equals 
\begin{equation} \label{specialko}
(L_{g_1}L_{g_2}R_{g_2^{-1}}\hat{e}_i,L_{g_2}\hat{e}_i)=:(\hat{\ce}_i^1,\hat{\ce}_i^2).
\end{equation}
Put $\hat{\ce}_i^+=(\hat{\ce}_i^1,\hat{\ce}_i^2)$ and $\hat{\ce}_i^-=(\hat{\ce}_i^1,-\hat{\ce}_i^2 )$. Since $\{ \hat{\ce}_i^\pm \}_i$ is an orthonormal frame for $TG^2$ it is natural to try to define
$$D_1(\xi )=\sum_i \ce_i^+\nabla_{\hat{\ce}_i^+}\xi+\sum_i \ce_i^-\nabla_{\hat{\ce}_i^-}\xi.$$
where $\xi\in L^2(G^2,Cl(T^*G^2))$ and where $\{\ce^\pm_i\}$ is the corresponding orthonormal frame for $T^*G^2$.
A small computation shows that 
$$\begin{array}{lc}
\hat{\ce}_i^+(P^*_{0,1}\xi)=2 P_{0,1}^*(\hat{E}_i(\xi)), & \xi\in L^2(G)\\ 
\hat{\ce}_i^-(P^*_{0,1}\xi)=0, &\xi\in L^2(G).
 \end{array}$$
We therefore define 
$$D_1(\xi )=\frac{1}{2} \left(\sum_i \ce_i^+\nabla_{\hat{\ce}_i^+}\xi+\sum_i \ce_i^-\nabla_{\hat{\ce}_i^-}\xi \right) .$$
Next, write an element in $\xi \in L^2(G,T^*G)$ as 
$$\xi =\sum_i \xi_i E_i,$$
where $\xi_i \in L^2(G)$. Hence 
$$P^*_{0,1}(\xi )=\sum_i P_{0,1}^*(\xi_i)\ce_i^+.$$ 
We calculate

\begin{eqnarray*}
\lefteqn{D_1(P^*_{0,1}(\xi ))}\\
&=&\sum_{i,j}\left( P_{0,1}^*(\hat{E}_i(\xi_j))\ce_i^+ \ce^+_j+\frac{1}{2}\left( P_{0,1}^*(\xi_j)\ce_i^+\nabla_{\hat{\ce}_i^+}\ce_j^++P_{0,1}^*(\xi_j)\ce_i^-\nabla_{\hat{\ce}_i^-}\ce_j^+\right)  \right).
\end{eqnarray*}
If we therefore require $\nabla$ to be a $SO(2\hbox{dim}(G))$-valued connection in $T^*G^2$ satisfying
$$\begin{array}{ccl}
\nabla_{\hat{\ce}_i^+}(\ce_j^+)&=&P^*_{0,1}(\nabla_{\hat{E}_i}(E_j))\\
 \nabla_{\hat{\ce}_i^-}(\ce_j^+)&=&0
\end{array}$$
we see that $P^*_{0,1}(D_0(\xi))=D_1(P^*_{0,1}(\xi))$ for all $\xi \in L^2(G,Cl(T^*G))$.

\subsection{A Riemannian metric }\label{riem}

Due to lemma \ref{gn} we have at each level the identification $\ca_k=G^{n_k}$, where $n_k$ is the number of edges in $\Gamma_k$. We want to construct a Riemannian metric on $T\ca_k =TG^{n_k}$. However, we require the metric to be consistent with the embeddings of graphs. Embeddings of graphs $\Gamma_k\subset \Gamma_{k+1}$ induces a surjective smooth map 
$$P_{k,k+1}: \ca_{k+1} \to \ca_{k},$$ 
and therefore a map of tangent bundles
$$(P_{k,k+1})_*:T\ca_{k+1}\to T\ca_{k}.$$
Dualizing this map we get an embedding
$$P_{k,k+1}^* :T^*\ca_{k} \to T^*\ca_{k+1}$$
of cotangent bundles. 

We want to construct the Hilbert space on which the Dirac operator acts as a inductive limit of Hilbert spaces. It is hence natural to construct the metric on the cotangent bundle of $\ca_k$, since we have canonical maps 
$$P_{k,k+1}^*:L^2(\ca_{k},T^*\ca_{k})\to L^2(\ca_{k+1},T^*\ca_{k+1}).$$

\begin{definition}
Let $\langle \cdot,\cdot \rangle$ be a left and right invariant Riemannian metric on $T^*G$. Let $\Gamma_k$ be the graph consisting of the edges in $\ct_k$, the $k$'s barycentric subdivision of $\ct_0$. On $T^*\ca_{k}=T^*G^{n_{k}}$ define
\begin{equation} \label{indre}
\langle \cdot ,\cdot \rangle_{k}=\frac{1}{2^k} \langle \cdot ,\cdot \rangle^{n_{k}},
\end{equation}
where  $\langle \cdot ,\cdot \rangle^{n_{k}}$ is the product metric on $T^*G^{n_{k}}$. 
\end{definition}

\begin{proposition}\label{komind}
The map 
$$P_{k, k+1}^*:T^*\ca_{k}\to T^*\ca_{k+1}$$
preserves the metric (\ref{indre}).
\end{proposition}

\textit{Proof.} We subdivide $G^{n_{{k+1}}}$ as $G^{2n_{k}}\times G^{n_2}$, where $G^{2n_{k}}$ corresponds to subdivision of the edges in $\Gamma_k$ and $G^{n_2}$ corresponds to the new edges added when going from $k$ to $k+1$. Write 
$$T^*_{(g_1,\ldots, g_{n_{k+1})}}\ca_{k+1}=T^*_{g_1}G\oplus \cdots \oplus T^*_{g_{n_{k+1}}}G.$$
We choose orientations of the edges in such a way that 
$$P_{k,k+1}(g_1,\ldots, g_{n_{k+1}})=(g_1g_2,g_3g_4,\ldots ,g_{2n_{k-1}}g_{2n_{k}}).$$
For a tangent vector ${\bf{v}}=(v_1,\ldots,v_{n_{k+1}})$ in $T^*\ca_{k+1}$ we have
$$(P_{k,k+1})_*({\bf{v}})=(L_{g_1}v_2+R_{g_2}v_1,L_{g_3}v_4+R_{g_4}v_3,\ldots ,L_{g_{2n_{k}-1}}v_{2n_{k}}+R_{g_{2n_{k}}}v_{2n_{k}-1}), $$
where $L_g$ means left translation of tangent vectors and $R_g$ right translation. Hence for a cotangent vector ${\bf{w}}=(w_1,\ldots, w_{n_{k}})$ in $T^*_{(g_1g_2, \ldots,g_{2n_{k}-1}g_{2n_{k}})}\ca_{k}$ we have
$$P^*_{k,k+1}({\bf{w}})=(R_{g_2^{-1}}w_1,L_{g_1^{-1}}w_1,\ldots,R_{g_{2n_{k}}^{-1}}w_{n_{k}},L_{2n_{k}-1}w_{n_{k}},0,\ldots,0),$$
where by definition 
$$L_g(w)(v)=w(L_{g^{-1}}(v)),\quad R_g(w)(v)=w(R_{g^{-1}}(v)), \quad v\in TG,\quad w\in T^*G.$$
From this and the left and right invariance of $\langle \cdot ,\cdot \rangle$ the proposition follows.\eproof \\

By proposition \ref{komind} the map $P_{k, k+1}^*$ induces a map, also denoted $P_{k ,k+1}^*$, from $Cl(T^*\ca_{k})$ to $Cl( T^*\ca_{k+1})$, where $Cl(T^*\ca_{k})$ denotes the Clifford bundle of $T^*\ca_{k}$ with respect to $\langle \cdot ,\cdot \rangle_{k}$. Furthermore this map is isometric. We thus get embeddings of Hilbert spaces
$$P^*_{k,k+1}:L^2(\ca_{k},Cl(T^*\ca_{k}))\to L^2(\ca_{k+1},Cl(T^*\ca_{k+1}).$$

\begin{definition}
Put $\ch_k=L^2(\ca_{k},Cl(T^*\ca_{k}))$. Define
$$\ch =\lim_{\longrightarrow }\ch_k,$$
the inductive limit of of the system $\{\ch_k, P_{k,k+1}^*\}_k$.
\end{definition}

Due to proposition \ref{induktiv} we can consider $\ch$ as $L^2(\overline{\ca}^\triangle ,Cl(\overline{\ca}^\triangle ))$ or more freely written $L^2(\ca,Cl(\ca))$.

\subsection{Some special covector fields}

For the general construction of the Dirac operator we will need some special covector fields generalizing duals of the vector fields (\ref{specialko}). For notational simplicity we will only present the construction on a single edge which is then subdivided infinitely many times. Thus 
$\ca_{n}=G^{2^n}$ and the structure maps are given by
$$P_{n-1,n} (g_1,g_2,\ldots,g_{2^n-1},g_{2^n})=(g_1g_2,\ldots, g_{2^n-1}g_{2^n}).$$
Let $\{ e_i\}$ be an orthonormal basis of $T_{id}^*G$. Define covector fields on $\ca_{0}=G$ by
$$\ce_i^{0,1}(g)=L_g(e_i).$$
The construction of the covector fields on $\ca_n$ will be by induction. Assume that  covector fields $$\{\ce_i^{j,k}\}_{j\leq n-1,k\leq 2^{j-1}}$$ on $\ca_{n-1}$ 
has been defined (For $j=0$ the set $k\leq 2^{j-1}$ is $\{ 1\}$). We adopt the notation
\begin{eqnarray*}
\ce^1_i &=&L_{g_1g_2\cdots g_{2^n}}R_{(g_2g_3\cdots g_{2^n})^{-1}}e_i\\
\ce^2_i &=&L_{g_2g_3\cdots g_{2^n}}R_{(g_3g_4\cdots g_{2^n})^{-1}}e_i\\
&\vdots
\end{eqnarray*}
for covector fields on $G$ depending on $g_1,\ldots ,g_{2^n}$.
Define covector fields on $\ca_{n}$ by
$$\{P_{n-1,n}^*(\ce_i^{j,k})\}_{j\leq n-1,k\leq 2^{j-1}}$$
and 
\begin{eqnarray*}
\ce^{n,1}_i &=&2^{n-1}(\ce_i^1,-\ce_i^2,0,\ldots,0)\\
\ce^{n,2}_i &=&2^{n-1}(0,0,\ce_i^3,-\ce_i^4,0,\ldots,0)\\
&\vdots
\end{eqnarray*}
With sloppy notation we will also write $\ce_i^{j,k}$ instead of $P_{n-1,n}^*(\ce_i^{j,k})$. 
\begin{lemma}
$\{\ce_i^{j,k}\}_{j\leq n,k\leq 2^{j-1}}$ is an orthonormal frame for $T^*\ca_{n}$ with respect to the inner product $\langle \cdot ,\cdot\rangle_{n}$.
\end{lemma}

\textit{Proof.} Since $P^*_{n-1,n}$ preserves the inner product by proposition \ref{komind} it is enough to check that 
$\{\ce_i^{n,k}\}_{k\leq 2^{j-1}}$ are orthonormal and are orthogonal to $\{\ce_i^{j,k}\}_{j\leq n-1,k\leq 2^{j-1}}$. 

It is clear that $\{\ce_i^{n,k}\}_{k\leq 2^{j-1}}$ are orthonormal. By induction and the computation in the proof of proposition \ref{komind}  we see that the vectors in $\{\ce_i^{j,k}\}_{j\leq n-1,k\leq 2^{j-1}}$ are of the form
$$(c_1\ce_i^1,c_1\ce_i^2,c_2\ce_i^3,c_2\ce_i^4,\ldots ,c_{2^{n-1}}\ce_i^{2^n-1},c_{2^{n-1}}\ce_i^{2^n}).$$ 
From this the lemma follows.\eproof

\subsection{Construction of the Dirac operator}
We want to construct a Dirac type operator acting on $\ch$. For this we construct  Dirac type operators acting on $\ch_n$ which are consistent with $P^*_{n,n+1}$.  We first need to determine sufficient conditions on the  connections which permit the existence of the Dirac operator.

Let $\hat{\ce}_i^{j,k}$ be the vector field obtained from $\ce_i^{j,k}$ by using $\langle \cdot ,\cdot \rangle_{n}$ to identify $T^*\ca_{n}$ with $T\ca_{n}$. Also let $\hat{e}_i$ denote the basis in $T_{id}G$ obtained from $e_i$ by using $\langle\cdot ,\cdot \rangle $ to identify $T_{id}G$ with $T^*_{id}G$. Note also  under this identification
the covectorfield
$$(0,\ldots ,0, \ce_i^j,0, \ldots) $$
gets mapped to
$$\frac{1}{2^{n}}(0,\ldots,0,L_{g_j\cdots g_{2^n}}R_{(g_{j+1}\cdots g_{2^n})^{-1}}\hat{e}_i,0 , \ldots,0).$$

\begin{definition}
A system of connections $\{ \nabla^n\}_n$, where $\nabla^n$ is a connection in $T^*\ca_{n}$ is called admissible if
$\nabla^n$ is a $SO(2^n\cdot \hbox{dim}(G))$-valued connection and if
$$\begin{array}{ccll}
\nabla^n_{\hat{\ce}_i^{j,k}}(\ce_l^{m,p})&=&P^*_{n-1,n}(\nabla^{n-1}_{\hat{\ce}_i^{j,k}}(\ce_l^{m,p}))&j,m<n\\
 \nabla^n_{\hat{\ce}_i^{n,k}}(\ce_l^{m,p})&=&0&m<n
\end{array}$$
\end{definition}
We want to define the Dirac operator in the usual fashion using an admissible family of connection, i.e. at each level it should be on the form
$$D_n=\sum \ce_i^{j,k}\cdot \nabla^n_{\hat{\ce}_i^{j,k}}.$$

\begin{proposition}\label{compat}
Let $\{ \nabla^n\}$ be an admissible system of connections and let $\{ a^{j,k}\}_{k\leq 2^{j-1}}$ be a sequence of complex numbers. For $\xi\in \ch_n$ define
$$D_n(\xi)=\sum_{j\leq n,k,i}a^{j,k}\ce_i^{j,k}\nabla^n_{\hat{\ce}_i^{j,k}}\xi.$$
Then 
$$P^*_{n,n+1}(D_n(\xi))=D_{n+1}(P^*_{n,n+1}(\xi))$$
and hence the system of operators $\{ D_n\}$ defines a densely defined operator $D$ on $\ch$.
\end{proposition}

\textit{Proof.} We first check the identity on functions, i.e. $\xi\in L^2(\ca_{n})$:
\begin{eqnarray*}
\lefteqn{P^*_{n,n+1}(D_n(\xi )) (g_1,g_2,\ldots ,g_{2^{n+1}})}\\
&=&P^*_{n,n+1}\left( \sum_{j\leq n,k,i}a^{j,k}\ce_i^{j,k}\cdot \hat{\ce}_i^{j,k}(\xi)\right) (g_1,\ldots ,g_{2^{n+1}})\\
&=&\sum_{j\leq n,k,i}a^{j,k}\ce_i^{j,k}\cdot P^*_{n,n+1}(\hat{\ce}_i^{j,k}(\xi) )(g_1,\ldots ,g_{2^{n+1}})\\
&=&\sum_{j\leq n,k,i}a^{j,k}\ce_i^{j,k}\cdot (\hat{\ce}_i^{j,k}(\xi) )(g_1g_2,\ldots ,g_{2^{n+1}-1}g_{2^{n+1}}).
\end{eqnarray*}
Let $\gamma_i$ be a curve in $G$ with $\dot{\gamma}_i(0)=\hat{e}_i$. Write
$$\ce_i^{j,k}=(c_1\ce_i^1,c_2\ce_i^2,\ldots ,c_{2^{n}}\ce_i^{2^n}).$$
Thus
\begin{eqnarray*}\lefteqn{\hat{\ce}_i^{j,k}(\xi) (g_1,\ldots ,g_{2^n})}\\
&=&\frac{1}{2^{n}}\sum_l c_l \frac{d}{dt}\xi(g_1,\ldots, g_{l-1},g_l\cdots g_{2^n}\gamma_i (g_{l+1}\cdots g_{2^n})^{-1},\\
&&g_{l+1},\ldots ,g_{2^n}),
\end{eqnarray*}
and
\begin{eqnarray*}\lefteqn{\hat{\ce}_i^{j,k}(\xi) (g_1g_2,\ldots ,g_{2^{n+1}-1}g_{2^{n+1}})}\\
&=&\frac{1}{2^{n}}\frac{d}{dt}\sum_l c_l \xi(g_1g_2,\ldots, g_{2(l-1)-1}g_{2(l-1)},g_{2l-1}\\
&&\cdots g_{2^{n+1}}\gamma_i (g_{2l+1}\cdots  g_{2^{n+1}})^{-1},g_{l+1},\ldots g_{2^{n+1}}).
\end{eqnarray*}
On the other hand
\begin{eqnarray*}
\lefteqn{D_{n+1}(P^*_{n,n+1}(\xi))(g_1,\ldots ,g_{2^{n+1}})}\\
&=&  \sum_{j\leq n+1,k,i}a^{j,k}\ce_i^{j,k}\cdot \hat{\ce}_i^{j,k}(P^*_{n,n+1}(\xi ))(g_1,\ldots,g_{2^{n+1}})\\
&=&\sum_{j\leq n,k,i}a^{j,k}\ce_i^{j,k}\cdot \hat{\ce}_i^{j,k}(P^*_{n,n+1}(\xi ))(g_1,\ldots,g_{2^{n+1}})\\
&&+\sum_{j = n+1,k,i}a^{j,k}\ce_i^{j,k}\cdot \hat{\ce}_i^{j,k}(P^*_{n,n+1}(\xi ))(g_1,\ldots,g_{2^{n+1}})
\end{eqnarray*}
The terms in the last sum are
\begin{eqnarray*} 
\lefteqn{\hat{\ce}_i^{n+1,k}(P^*_{n,n+1}(\xi ))(g_1,\ldots,g_{2^{n+1}})}\\
&=& \frac{1}{2}\frac{d}{dt} \big( \xi (g_1g_2,\ldots,g_{2k-1}\cdots g_{2^{n+1}}\gamma_i(g_{2k}\cdots g_{2^{n+1}})^{-1}g_{2k},\ldots,g_{2^{n+1}-1}g_{2^{n+1}})\\
&& -\xi (g_1g_2,\ldots,g_{2k-1} g_{2k}\cdots g_{2^{n+1}}\gamma_i(g_{2k+1}\cdots g_{2^{n+1}})^{-1},\\
&&g_{2k+1},\ldots,g_{2^{n+1}-1}g_{2^{n+1}})\big) \\
&=&0
\end{eqnarray*}
The terms in the first sum are
\begin{eqnarray*}
\lefteqn{\hat{\ce}_i^{j,k}(P^*_{n,n+1}(\xi ))(g_1,\ldots,g_{2^{n+1}})}\\
&=&\frac{1}{2^{n+1}}\frac{d}{dt}\sum_l c_l\big( \xi (g_1g_2,\ldots ,\\
&&g_{2l-1}\cdots g_{2^{n+1}}\gamma_i(g_{2l}\cdots g_{2^{n+1}})^{-1}g_{2l},\ldots,g_{2^{n+1}-1}g_{2^{n+1}})\\
&&+\xi (g_1g_2,\ldots , g_{2l-1}g_{2l}\cdots g_{2^{n+1}}\gamma_i(g_{2l+1}\cdots g_{2^{n+1}})^{-1},\ldots,g_{2^{n+1}-1}g_{2^{n+1}})\big)\\
&=&\frac{1}{2^{n}}\frac{d}{dt}\sum_l c_l \xi(g_1g_2,\ldots, g_{2(l-1)-1}g_{2(l-1)},g_{2l-1}\\
&&\cdots g_{2^{n+1}}\gamma_i (g_{2l+1}\cdots  g_{2^{n+1}})^{-1},g_{l+1},\ldots g_{2^{n+1}}).
\end{eqnarray*}
This proves the compatibility for functions.

Because of the derivation property of the connection on the Clifford bundle it only remains to prove
compatibility on vectors of the form $\ce_i^{j,k}$, $j\leq n$. This follows from
\begin{eqnarray*}
\lefteqn{P^*_{n,n+1}(D_n( \ce_l^{m,p}))}\\
&=&P^*_{n,n+1}(\sum_{j\leq n,k,i}a^{j,k}\ce_i^{j,k}\nabla^n_{\hat{\ce}_i^{j,k}}\ce_l^{m,p})\\
&=&\sum_{j\leq n+1,k,i}a^{j,k}\ce_i^{j,k}\nabla^{n+1}_{\hat{\ce}_i^{j,k}}(\ce_l^{m,p})\\
&=&D_{n+1}(\ce_l^{m,p})
\end{eqnarray*}
where we have used the admissibility condition for the connections.\eproof

\subsection{Gauge invariance}
In section \ref{gauge} we have constructed a left action of $\overline{\cg}^\triangle$ on $\overline{\ca}^\triangle$. It follows from the construction that this action preserves the inner product on $L^2(\overline{\ca}^\triangle)$, and hence is a unitary action of $\overline{\cg}^\triangle$ on $L^2(\overline{\ca}^\triangle)$. In order to talk about gauge invariance of of the Dirac operator we first need to extend this action to a unitary action on $L^2(\overline{\ca}^\triangle, Cl(\overline{\ca}^\triangle))$. 

Let $U$ be a gauge transformation written as 
$$U_n (g_1,\ldots, g_{2^n})=(u_0g_1u_1^{-1},\ldots , u_{2^n-1}g_{2^n}u_{2^n}^{-1}),$$
on $\ca_n$, where $u_0,\ldots , u_{2^n}\in G$. Since the metric is left and right invariant in each copy of $G$, we see that $(U_n)^*:T^*\ca_n\to T^*\ca_n$ preserves the metric and therefore $U$ extends to a unitary on $L^2(\overline{\ca}^\triangle, Cl(\overline{\ca}^\triangle))$. This unitary will also be denoted by $U$. 

A vectorfield of the form 
$$(g_1,\ldots, g_{2^n})\to (0,\ldots,0,L_{g_j\cdots g_{2^n}}R_{(g_{j+1}\cdots g_{2^n})^{-1}}\hat{e}_i,0 , \ldots,0)$$
gets mapped to
$$  (u_0g_1u_1^{-1},\ldots , u_{2^n-1}g_{2^n}u_{2^n}^{-1}) \to (0,\ldots,0,L_{u_{j-1}g_j\cdots g_{2^n}}R_{(u_{j}^{-1}g_{j+1}\cdots g_{2^n})^{-1}}\hat{e}_i,0 , \ldots,0),$$
under $(U_n)_*$. Written differently
$$(g_1,\ldots, g_{2^n})\to (0,\ldots,0,L_{g_j\cdots g_{2^n}}R_{(g_{j+1}\cdots g_{2^n})^{-1}}u_{2^n}\hat{e}_iu_{2^n}^{-1},0 , \ldots,0).$$
Therefore $(U_n)^*$ is, on a covector field of the form 
$$(g_1,\ldots, g_{2^n})\to (0,\ldots,0,L_{g_j\cdots g_{2^n}}R_{(g_{j+1}\cdots g_{2^n})^{-1}} e_i,0 , \ldots,0),$$
given by
$$(g_1,\ldots, g_{2^n})\to (0,\ldots,0,L_{g_j\cdots g_{2^n}}R_{(g_{j+1}\cdots g_{2^n})^{-1}}u_{2^n}^{-1}e_iu_{2^n},0 , \ldots,0).$$
Put $f_i=u_{2^n}e_iu_{2^n}$. Since $\{f_i\}$ be another orthonormal basis for the dual of the Lie algebra of $G$, there exist a matrix $O_{ij} \in O(dim(G))$ with 
$$f_i=\sum_j O_{ij}e_j.$$ 
 From the construction of the covector fields  $\{\ce_i^{j,k}\}$ it follows that 
$$U(\ce_i^{j,k})=\sum_l O_{il}\ce_l^{j,k}.$$

It turns out that the operator $D$ is not always gauge invariant. We therefore need
\begin{definition}
 An admissible system of connections is gauge admissible if the conditions 
$$\sum_{r,i,q} O_{ql} O_{ir}\ce_r^{j,k}U( \nabla^n_{\hat{\ce}_i^{j,k}} \ce_q^{m,p})=\sum_i\ce_i^{j,k}\nabla^n_{\hat{\ce}_i^{j,k}}\ce_l^{m,p}$$
hold for all $n,j,k$ and all gauge transformations $U\in \overline{\cg}^\triangle$.
\end{definition}

Note that the system of trivial connections with respect to the trivializations given by $\{\ce_i^{j,k}\}$ is a gauge admissible system of connections.

\begin{prop} \label{megetgrim}
When $D$ is constructed from a gauge admissible connection,  $D$ is gauge invariant, i.e. $D=UDU^*$ for all $U\in \overline{\cg}^\triangle$.
\end{prop}

\textit{Proof.} 
Let $\xi \in L^2(\ca_n)$. We compute
\begin{eqnarray*}
 \lefteqn{UDU^*(\xi)}\\
&=&\sum_{j\leq n,k,i}a^{j,k}(U_n)^*(\ce_i^{j,k}) d_{(U_n^{-1})_*(\hat{\ce}_i^{j,k})}\xi\\
&=& \sum_{j\leq n,k}a^{j,k}\sum_{i,l,m}O_{il}\ce_l^{j,k} d_{O_{im}\hat{\ce}_m^{j,k}}\xi \\
&=& \sum_{j\leq n,k,i}a^{j,k}\ce_i^{j,k} d_{\hat{\ce}_i^{j,k}}\xi \\
&=& D(\xi)
\end{eqnarray*}

Due to the derivation and product structures we now only need to check gauge invariance on the covectorfields 
$\{\ce_i^{j,k}\}$. Here we get
\begin{eqnarray*}
 \lefteqn{UDU^*(\ce_l^{m,p})}\\
&=& U (\sum_{j\leq n,k,i,q}a^{j,k}\ce_i^{j,k}O_{ql} \nabla^n_{\hat{\ce}_i^{j,k}} \ce_q^{m,p})\\
&=& \sum_{j\leq n,k,i,q}a^{j,k}U(\ce_i^{j,k}O_{ql}) U( \nabla^n_{\hat{\ce}_i^{j,k}} \ce_q^{m,p})\\
&=&\sum_{j\leq n,k}a^{j,k}\left(\sum_{r,i,q} O_{ql} O_{ir}\ce_r^{j,k}U( \nabla^n_{\hat{\ce}_i^{j,k}} \ce_q^{m,p})\right)\\
&=&\sum_{j\leq n,k,i}a^{j,k}\ce_i^{j,k}\nabla^n_{\hat{\ce}_i^{j,k}}\ce_l^{m,p}\\
&=&D(\ce_l^{m,p}),
\end{eqnarray*}
where we have used gauge admissibility of $\nabla^n$. \eproof

\section{A semifinite spectral triple} \label{semi}
In this section we will construct a semifinite spectral associated to $\overline{\ca}^\triangle$ or rather to the algebra of holonomy loops, see definition \ref{holalg}. The following definition first appeared \cite{CPS}.
\begin{definition}
Let $\cn$ be a semifinite von Neumann algebra with a semifinite trace $\tau$. Let $\ck_\tau$ be the $\tau$- compact operators. A semifinite spectral triple $(\cb ,\ch,D)$ is a $*$-subalgebra $\cb$ of $\cn$, a representation of $\cn$ on the Hilbert space $\ch$ and an unbounded densely defined self adjoint operator $D$ on $\ch$ affiliated with $\cn$ satisfying
\begin{enumerate}
\item $b (\lambda -D)^{-1}\in \ck_\tau$ for all $b\in \cb$ and $\lambda \notin \bbR$.. 
\item $[ b,D ]$ is densely defined and extends to a bounded operator.
\end{enumerate}
\end{definition}

\subsection{The algebra}
Let $v$ be a vertex in $\cs_\triangle$. Denote by $\ch\cg_v(S_\triangle )$ the subgroupoid of $\ch\cg (S_\triangle )$ of loops based in $v$, i.e. paths starting and ending in $v$. Let $G\to M_N$ be a unitary matrix representation of $G$ and let $\ch'=\ch \otimes M_N$. Hence 
$$\ch'= \lim_{\longrightarrow} (\ch_n \otimes M_N).$$
\begin{definition}\label{holalg}
Let $L$ be a loop in $\ch\cg_v(S_\triangle )$. Let $\xi \in \ch_n\otimes M_N$, where $n$ is such that $L\in \ch\cg (n )$. Define 
$$L( \xi )(\nabla )=\nabla (L) \xi (\nabla), \quad \nabla \in \ca_{n}=Hom(\ch \cg ( n),G).$$
Since 
$$P^*_{n,m}(L(\xi))=L(P^*_{n,m}(\xi))$$ 
we get a densely defined operator on $\ch'$, also denoted $L$. Clearly $L$ is bounded.

Denote by $\cb_v$ the $*$-algebra generated by $\{ L\}_{L\in  \ch\cg_v(S_\triangle )}$. 
We will call $\cb_v$  the algebra of holonomy loops.
\end{definition}

\begin{proposition}
For $L\in  \ch\cg_v(S_\triangle )$ define the function
$HL:\ca\to M_N $ by
$$HL(\nabla )=Hol(L,\nabla).$$
The $*$-algebra generated by $\{HL\}_{L\in  \ch\cg_v(S_\triangle )}$ equipped with the sup norm is isomorphic to $\cb_v$ as normed $*$-algebra. 
\end{proposition}

\textit{Proof.} Follows from the dense embedding $\ca\to \overline{\ca}^\triangle$. \eproof

\subsection{The trace and the von Neumann algebra} \label{von}
Let
$$A_n=\ck(L^2(\ca_{n}))\otimes Cl(T^*_{id}\ca_{n})\otimes End(M_N).$$
where $\ck(L^2(\ca_{n}))$ denotes the compact operators on $L^2(\ca_{n})$. 
We define maps, with an abuse of notation also denoted $P^*_{n,n+1}$, from $A_n$ to $A_{n+1}$ in the following way
\begin{itemize}
\item On $End(M_N)$, $P_{n,n+1}^*$ is the identity.
\item On $\ck (L^2(\ca_{n}))$, $P_{n,n+1}^*$ is the map induced by the embedding $$P_{n,n+1}^*:L^2 (\ca_{n})\to L^2(\ca_{n+1}),$$
i.e. if $P_{n,n+1}$ denotes the projection in $L^2(\ca_{n+1})$ onto $P_{n,n+1}^*(L^2(\ca_{n}))$ we have
$$\ck (L^2(\ca_{n}))\ni a\to aP_{n,n+1}\in \ck (L^2(\ca_{n+1})).$$
\item We  use the map 
$$P_{n,n+1}:Cl(T^*\ca_n) \to Cl(T^*\ca_{n+1}),$$
which has already been defined in subsection \ref{riem} and restrict it to the identity. If we write 
$$T^*_{id}\ca_{n+1}=T^*_{id}\ca_{n}\oplus V_{n,n+1},$$
and hence  
$$Cl(T^*_{id}\ca_{n+1})=Cl(T^*_{id}\ca_{n})\hat{\otimes} Cl( V_{n,n+1}),$$
then $$P^*_{n,n+1}:Cl(T^*_{id}\ca_{n})\to Cl(T^*_{id}\ca_{n+1})$$ 
is given by
$$P^*_{n,n+1}(\ce)=\ce\otimes \mathbf{1}_{Cl( V_{n,n+1})}.$$
\end{itemize}
Note that $P^*_{n,n+1}$ is a $C^*$-algebra homomorphism. Therefore $\{ A_n\}$ is an inductive system of $C^*$-algebras. Let 
$$A=\lim A_n$$ 
be the inductive limit.  By 
$$P^*_{n,\infty} :A_n \to A,$$
we denote the induced embeddings.  

From the construction of $A$ it follows that 
\begin{equation}\label{faktor}
A=\ck(L^2(\ca))\otimes B \otimes M_N,
 \end{equation}
 where $B$ is a UHF-algebra.  Since the dimension of the Clifford algebra is a power of $2$ when $n\geq1 $, $B$ is  the CAR-algebra.

There is a trace $Tr$ on $A_k$ defined by
$$Tr=Tr_o\otimes Tr_n \otimes Tr_o,$$
where $Tr_o$ are the operator traces on $\ck(L^2(\ca_{k}))$, resp. $ End(M_N)$ and $Tr_n$ is the normalized trace on $ Cl(T^*_{id}\ca_{k})$. By construction  
$$Tr ( a )=Tr(P^*_{k,k+1}(a )),$$ and hence defines a densely defined trace on $A$.

Using  the trivialization $\{\ce_i^{j,k}\}$ of $T^*_{id}\ca_{n}$ we see that  $A$ acts on $\ch'$. In fact $\ch'$ factors like (\ref{faktor}) into
$$\ch'=L^2(\ca )\otimes Cl(T^*_{id}\ca)\otimes M_N.$$
Note that the action of $B$ on $Cl(T^*_{id}\ca)$ is just the GNS-representation of $B$ with respect to the normalized trace on $B$.
\begin{definition}
Let $\cn$ be the weak closure of $A$ in $\cb (\ch')$. The trace $\tau :\cn \to \bbC$ is defined as the extension of the trace $Tr$ on $A$ to $\cn$. 
\end{definition}

Note that $\tau$ is a semifinite trace, since it is the tensor product of the usual semifinite trace on $\cb(\ch)$, $\ch$ separable, and the finite trace on the hyperfinite $\hbox{II}_1$ factor $\overline{B}^w$. 

\subsection{A coordinate change and selfadjointness of the Dirac operator} 
We define the coordinate transformation 
$$\Theta_{n}:\ca_{n}=G^{2^n}\to G^{2^n}$$
by 
$$\Theta_{n} (g_1,\ldots, g_{2^n})=(g_1g_2\cdots g_{2^n},g_2g_3\cdots g_{2^n},\ldots , g_{2^n-1}g_{2^n} , g_{2^n}).$$
It is easy to see that $\Theta_{n}$ preserves the Haar measure on $G^{2^n}$.

The inverse of $\theta_n$ is given by
$$\Theta_{n}^{-1}(g_1,\ldots ,g_{2^n})=(g_1g_2^{-1},g_2g_3^{-1},\ldots, g_{2^n-1}g_{2^{n}}^{-1},g_{2^{n}}),$$

The main purpose of $\Theta_{n}$ is the following:

$\ca_{n}$ is a trivial $G^{2^{n-1}}$-principal fiber bundle over $\ca_{n-1}$, where the action of $G^{2^{n-1}}$ on $\ca_{n}$ is given by
$$(g_1,\ldots ,g_{2^{n-1}})(g'_1,\ldots,g'_{2^n})=(g'_1g_1^{-1},g_1g'_2,\ldots , g'_{2^n-1}g_{2^{n-1}}^{-1},g_{2^{n-1}}g'_{2^n}).$$
Combining this with $\Theta_{n}$ we get the following commutative diagram
\begin{displaymath}
\begin{array}{ccccl}
&&\vdots&&\vdots\\
G^{2^{n}}&\to & \ca_{n}&\stackrel{\Theta_{n}}{\rightarrow}&G^{2^{n}} \\
&&\downarrow&&\downarrow pr_o\\
G^{2^{n-1}}&\to &\ca_{n-1}&\stackrel{\Theta_{n-1}}{\rightarrow}&G^{2^{n-1}}\\
&&\vdots&&\vdots\\
G&\to &\ca_{1}&\stackrel{\Theta_1}{\rightarrow}&G^2\\
&&\downarrow&&\downarrow pr_o\\
&&\ca_{0}&=&G
\end{array}
\end{displaymath}
where $pr_o$ means projection onto the odd coordinates, i.e. 
$$pr_o(g_1,g_2,\ldots,g_{2^n-1},g_{2^n})=(g_1,g_3,\ldots ,g_{2^n-1}).$$ 
In other words $\{ \Theta_n\}$ is just a consistent way of trivializing the principal bundles. 

\begin{lemma}  \label{grimt}
Let $ \hat{E}_i (g)=L_g(\hat{e}_i)$, the left translate of $\hat{e}_i$. On $\ca_{n}$ we have
$$((\Theta_{n})_* (\hat{\ce}_i^{n,k}))(g_1,\ldots ,g_{2^n})=-\frac{1}{2 }(0,\ldots,0, \hat{E}_i(g_{2k}),0,\ldots,0),$$
and if we write 
$$\hat{\ce}^{j,k}_i=(c_1\hat{\ce}_i^1,c_1\hat{\ce}_i^2,c_2\hat{\ce}_i^3,c_2\hat{\ce}_i^4,\ldots,c_{2^{n-1}}\hat{\ce}_i^{2^n-1},c_{2^{n-1}}\hat{\ce}_i^{2^n}), \quad j < n,$$
then
\begin{eqnarray*} &&\hspace{-2cm}
((\Theta_{n})_* (\hat{\ce}^{j,k}_i))(g_1,\ldots ,g_{2^n})=\\
&&
\Big( 2\sum_{j=1}c_j\hat{E}_i(g_1),(c_1+2\sum_{j=2}c_j)\hat{E}_i(g_2),2\sum_{j=2}c_j\hat{E}_i(g_3),\ldots ,\\
&& 2c_{2^{n-1}}\hat{E}_i(g_{2^n-1}),c_{2^{n-1}}\hat{E}_i(g_{2^n}) \Big)
\end{eqnarray*}
\end{lemma}

\textit{Proof.} Straightforward computation. \eproof\\

We will also write $\hat{E}_i^j(g_1,\ldots,g_{2^n})=(0,\ldots, \hat{E}_i(g_j),\ldots,0)$.

\begin{proposition}\label{selv}
Let $\{\nabla^n\}$ be an admissible system of connections. Assume  that $D_0$ is self adjoint, 
$$\nabla^n_{\hat{\ce}_i^{j,k}} (\ce_l^{n,m})=0  \hbox{ for all }i,j,k,m,l,\hbox{ and  all } n\geq 1 $$
and that $a^{j,k}$ are real and non zero for all $j,k$.
Then $D$ is self adjoint.
\end{proposition}

\textit{Proof.} We will prove that $D_n$ is formally self adjoint for each $n$. From this the statement follows because $D_n$ is an elliptic pseudo differential operator on a compact manifold and therefore by elliptic regularity $D_n$ is self adjoint. We can even find an orthonormal basis for $\ch'_n$ which diagonalizes $D_n$ with real eigenvalues. We can therefore find an orthonormal basis for $\ch'$ which diagonalizes $D$ with real eigenvalues. Hence $D$ is self adjoint.

Write
$$D_n=\sum_{i}a^{0,1}\ce_i^{0,1}\nabla^n_{\hat{\ce}_i^{0,1}}+\sum_{1\leq j\leq n,k,i}a^{j,k}\ce_i^{j,k}\nabla^n_{\hat{\ce}_i^{j,k}}.$$ 
By the assumptions on the connection and lemma \ref{grimt} the first summand on the right hand side is a linear combination of $D_0$'s acting on the different copies of $G$, and thus by assumption self adjoint. The second summand is, in the trivialization induced by $\{\ce_i^{j,k} \}$ and transported by $\Theta_n$ on the form
$$ \sum_{1\leq j\leq n,k,i}a^{j,k}\ce_i^{j,k} d_{(\Theta_n)_*(\hat{\ce}_i^{j,k})},$$ 
where $d$ is the exterior derivative. Since  $(\Theta_n)_*(\hat{\ce}_i^{j,k})$ is a left invariant vector field according to lemma \ref{grimt}, $d_{(\Theta_n)_*(\hat{\ce}_i^{j,k})}$ is formally skew self adjoint. The formal selfadjointness now follows since Clifford multiplication with $ \ce_i^{j,k}$ is also skew self adjoint and commutes with $d_{(\Theta_n)_*(\hat{\ce}_i^{j,k})}$. \eproof 

\subsection{Affiliation of $D$}
The spectral projections of $D_n$ will by construction belong to 
$$\ck(L^2(\ca_{n}))\otimes End (Cl(T^*_{id}\ca_{n})) \otimes End(M_N).$$
Since we can split $Cl(T^*_{id}\ca_{n})$ into irreducible representations of  $Cl(T^*_{id}\ca_{n})$, and $D_n$ acts on each of these,  the spectral projections of $D_n$ is in 
$$A_n=\ck(L^2(\ca_{n}))\otimes Cl(T^*_{id}\ca_{n}) \otimes End(M_N).$$

Recall from section \ref{von} the splitting
$$T^*_{id}\ca_{n+1}=T^*_{id}\ca_{n}\oplus V_{n,n+1}.$$
More generally we can write
$$T^*_{id}\ca_{m}=T^*_{id}\ca_{n}\oplus V_{n,m},$$
and we thus have
\begin{equation} \label{tensor}
Cl(T^*_{id}G^m)=Cl(V_{n,m})\hat{\otimes} Cl(T^*_{id}\ca_{n}).
\end{equation}

We assume that the system of connections defining the Dirac operator satisfies the properties in \ref{selv}.
These properties ensure the following equation:
$$D_m(P^*_{n,m}(\xi)\otimes v)= P^*_{n,m}(D_n(\xi))\otimes v, \quad \xi\in \ch'_n, v\in Cl(V_{n,m}).$$
If $\xi$ is an eigenvector with eigenvalue $\lambda$, then $P^*_{n,m}(\xi )\otimes v$ is an eigenvector with eigenvalue $\lambda$ if $v\in Cl(V_{n,m})$. Therefore, if $P_{\lambda, n}$ is the spectral of eigenvalue $\lambda$ of $D_n$ the projection $P_{n,\infty}^*(P_{\lambda ,n})$ is a subprojection of the spectral projection of eigenvalue $\lambda$ of $D$. By construction $P_{n,\infty}^* (P_{\lambda ,n}) \in \cn$

\eproof
\begin{prop}
The spectral projections of the Dirac operator is contained in $\cn$.
\end{prop}

\textit{Proof.} For a spectral projection $P_\lambda$ of eigenvalue $\lambda$ of $D$ we have 
$$P_{n,\infty}^*(P_{\lambda ,n}) \nearrow P_\lambda .$$
Hence $P_\lambda$ is in the weak closure of $A$, i.e. $P_\lambda \in \cn$. \eproof 

\begin{cor}\label{affil}
The operator $D$ is affiliated with $\cn$. 
\end{cor}

\subsection{The main theorem}
We begin by enlarging our Hilbert space slightly. Let $\ch=\ch'\otimes Cl(1)$. We also enlarge $\cn$ by tensoring with $Cl(1)$. By abuse of notation we will also call the enlargement $\cn$. The trace $\tau$ is also enlarged by tensoring with the normalized trace on $Cl(1)$ and denoted by $\tau$.

The Dirac operator  extends to an operator on $\ch$. We write
$$D_n= \sum_{j < n,k,i}a^{j,k}\ce_i^{j,k}\nabla^n_{\hat{\ce}_i^{j,k}}+\sum_{k,i}a^{n,k}\ce_i^{n,k}\nabla^n_{\hat{\ce}_i^{n,k}}=D_n^++D_n^-.$$ 
After changing coordinates with $\Theta_n$ we have 
$$D_n^-=\sum_{k,i}a^{n,k}\ce_i^{n,k}\nabla^n_{\hat{E}_i^{2k}}.$$
Therefore, when we use $\Theta_n$ to trivialize 
$$\ca_{n}=\ca_{n-1}\times G^{2^{n-1}}$$ we see that $D_n^-$ is an operator acting  only in the fiber, i.e. in the $G^{2^{n-1}}$ part of $\ca_{n}$. There is an embedding of 
\begin{equation} 
 L^2(\ca_{n-1})\otimes Cl(T^*_{id}\ca_{n})\otimes M_N\otimes Cl(1)
\label{linie3}
\end{equation}
 in $Ker (D^-_n)$ by identifying it as a subspace of $\ch_n$. Let $P_n$ be the projection onto the orthogonal complement of (\ref{linie3}) in $Ker (D^-_n)$. Define
$$D_{n,p}=D_n+\sum_j b^jeP_j,$$
where $b_j\in i \bbR$, $e$ is the generator of $Cl(1)$ and where we have used $P_{j,{j+1}}^*$ to push forward $P_j$. 
Note that $D_{n,p}$ is self adjoint. By construction 
$$P_{n,n+1}^*(D_{n,p}(\xi))=D_{n+1,p} (\xi),$$
and therefore $\{ D_{n,p}\}$ defines a densely defined self adjoint operator $D_p$ on $\ch$. Since $D$ is affiliated with $\cn$ by corollary \ref{affil}, so is $D_p$ by construction.   
\begin{thm} \label{hoved}
There exist sequences $\{ a^{j,k} \}$ and $\{ b_k\}$ such that $(\cb_v ,\ch  ,D_p )$ is a semifinite spectral triple with respect to $(\cn ,\tau)$.
\end{thm}

\textit{Proof.} The selfadjointness and affiliation to $\cn$ of $D_p$ are already taken care of. What remains is to prove that $[ b, D_p]\in \cb (\ch )$ and that $b (\lambda -D_p)^{-1}\in \ck_\tau$ for all $b\in \cb_v$. 

The boundedness of the commutator: Let $L$ be a loop in $\Gamma_n$. If we write 
$$\ch_n=L^2(G^{2^n})\otimes Cl(T^*\ca_{n})\otimes M_N \otimes Cl(1)$$ the loop operator $L$ acts by point wise matrix multiplication over $G^{2^n}$ in the $M_N$ factor, i.e. the matrix entries is of the form $f(g_1,\ldots , g_{2^n})$. 
The action  of $L$ on $\ch_{n+1}$ is then matrix multiplication with entries of the form $f(g_1g_2,\ldots , g_{2^{n+1}-1}g_{2^{n+1}})$. 
Conjugating the operator with $\Theta_{n+1}$ we see, since 
$$\Theta_{n+1}^{-1}(g_1,\ldots ,g_{2^{n+1}})=(g_1g_2^{-1},g_2g_3^{-1},\ldots, g_{2^n-1}g_{2^{n+1}}^{-1},g_{2^{n+1}}),$$
that the result is independent of $g_2,g_4,\ldots ,g_{2^{n+1}}$, and thus $[L,D_{n+1}^- ]=0$. It therefore follows that $[L,P_{n+1}]=0$. Hence $[L,D_{n+1,p}]=[L,D_{n,p}]$ and therefore $[L,D_p]$ is bounded.

To prove that there exist sequences $\{a^{j,k} \}$ and $b_j$ such $D_p$ has $\tau$-compact resolvent we will prove that for any real sequence $c_n$ converging to $\infty$ we can choose $a^{n,k}$ and $b_n$ such that the new eigenvalues, modulo the extra multiplicity of the existing eigenvalue due to the growth of the Clifford bundle, introduced by going from $(D_{n-1,p})^2$ to $(D_{n,p})^2$ are bigger than $c_n$.   

In the following we will omit  the $M_N$ part. This will play no role, since the Dirac operator does not act on the $M_N$ part. First we rewrite the operator in the following way
$$D_{n,p}=\left( D^+_n+\sum_{j\leq n-1}b_jeP_j  \right) +\left( D^-_n+b_neP_n \right) =D_{n,p}^++\left( D^-_n+b_neP_n \right),$$
and 
\begin{eqnarray*}
\lefteqn{\left( D^-_n+b_neP_n \right)}\\
&=&\sum_{k,i}a^{n,k}\ce_i^{n,k}\nabla^n_{\hat{\ce}_i^{n,k}}+b_n eP_n
\\&=& a\left( \sum_{k,i}\frac{a^{n,k}}{a}\ce_i^{n,k}\nabla^n_{\hat{\ce}_i^{n,k}}+\frac{b_n}{a} eP_n \right)=:aD^-_{n,p}.
\end{eqnarray*}
Using the coordinate change $\Theta_n$ we factorize 
\begin{eqnarray*}\ch_n&=& L^2(G^{2^{n-1}})\otimes L^2(G^{2^{n-1}})\otimes Cl(T^*_{id}\ca_{n})\otimes Cl(1)\\
&=&\ch^1\otimes \ch^2 \otimes Cl(T^*_{id}\ca_{n})\otimes Cl(1),
\end{eqnarray*}
where $\ch^1=L^2(G^{2^{n-1}})$ corresponds to the coordinates $(g_1,g_3,\ldots , g_{2^n-1})$ under $\Theta_n$ and $\ch^2$ corresponds to the coordinates $(g_2,g_4,\ldots , g_{2^n})$. In particular by lemma \ref{grimt}, $D_{n,p}^-$ acts trivially on $\ch^1$. Taking the square of $D_{n,p}$ we get
\begin{equation}
(D_{n,p})^2=(D_{n,p}^+)^2+a\{ D_{n,p}^- ,D_{n,p}^+ \}+a^2 (D_{n,p}^-)^2.
\label{eet}
\end{equation}
Using lemma \ref{grimt} it is easy to see that $ \{ D_{n,p}^- ,D_{n,p}^+ \}$ does not act on $\ch^1$. 

Let $\xi$ belong to the orthogonal complement of $\ch^1 \otimes 1 \otimes Cl(T^*_{id}\ca_{n}) \otimes Cl(1)$. We know, due to the fact that $D_p$ is self adjoint and commutes with the maps $P^*_{k,k+1}$ that  this complement is an invariant subspace for $D_p$.
Decompose $\xi$ with respect to the above decomposition of $\ch_n$ into 
\begin{equation}
\xi = \sum_k  \xi^1_k\otimes \xi_k^2 \;,
\label{too}
\end{equation}
where $\{\xi_k^1\}$ is an orthonormal basis for $\ch^1$ and $\xi_k^2$ belongs to the orthogonal complement of 
$$1 \otimes Cl(T^*_{id}\ca_{n}) \otimes Cl(1)$$ 
in 
$$\ch^2 \otimes Cl(T^*_{id}\ca_{n}) \otimes Cl(1).$$ Combining (\ref{eet}) and (\ref{too}) we get
\begin{eqnarray*}
\lefteqn{\langle (D_{n,p})^2 \xi,\xi\rangle} \\
&=& \langle  (D^+_{n,p})^2\xi,\xi\rangle +a\langle \{ D^+_{n,p},D^-_{n,p}\}\xi,\xi\rangle + a^2\langle (D^-_{n,p})^2\xi,\xi\rangle\\
&\geq & \langle \{ D^+_{n,p},D^-_{n,p}\}\xi,\xi\rangle + a^2 \langle (D^-_{n,p})^2\xi,\xi\rangle \\
&=&\sum_k \langle \xi_k^1,\xi_k^1 \rangle \left(a\langle \{ D^+_{n,p},D^-_{n,p}\}\xi_k^2 ,\xi_k^2 \rangle+
a^2\langle (D^-_{n,p})^2 \xi_k^2 ,\xi_k^2 \rangle \right)
\end{eqnarray*}
Since $(D^-_{n,p})^2$ is an elliptic second order operator on $\ch^2\otimes Cl(T^*_{id}\ca_{n}) \otimes Cl(1)$ and the operator $\{ D^+_{n,p},D^-_{n,p}\}$ is a first order operator  on $\ch^2 \otimes Cl(T^*_{id}\ca_{n}) \otimes Cl(1)$, the operator
$$A=(D^-_{n,p})^{-1}\{ D^+_{n,p},D^-_{n,p}\} (D^-_{n,p})^{-1}$$
is bounded.  We thus get
\begin{eqnarray*}
\lefteqn{\langle ((D_{n,p})^2 \xi,\xi\rangle} \\
&\geq&\sum_k \langle \xi_k^1,\xi_k^1 \rangle \left(a\langle \{ D^+_{n,p},D_{n,p}^-\}\xi_k^2 ,\xi_k^2 \rangle+
a^2\langle (D^-_{n,p})^2 \xi_k^2 ,\xi_k^2 \rangle \right)\\
&=&\sum_k \langle \xi_k^1,\xi_k^1 \rangle ( a\langle A D^-_{n,p}\xi_k^2 ,D^-_{n,p}\xi_k^2 \rangle
+a^2\langle (D^-_{n,p})^2 \xi_k^2 ,\xi_k^2 \rangle )\\
&\geq & \sum_k \langle \xi_k^1,\xi_k^1 \rangle \left( -a\|A\|\langle  D^-_{n,p}\xi_k^2 ,D^-_{n,p}\xi_k^2 \rangle
+a^2\langle (D^-_{n,p})^2 \xi_k^2 ,\xi_k^2 \rangle \right)\\
&\geq & \sum_k \langle \xi_k^1,\xi_k^1 \rangle \lambda (a^2-a\|A\|)\langle  \xi_k^2 ,\xi_k^2 \rangle\\
&=&\lambda (a^2-a\| A\| )\|\xi\|^2
\end{eqnarray*}
where $\lambda$ is the lowest eigenvalue of $(D_{n,p}^-)^2$ on the complement of 
$$\ch^1\otimes 1 \otimes Cl(T^*_{id}\ca_{n}) \otimes Cl(1).$$
Hence by choosing $a$ big enough  we have $\langle (D_{n,p})^2\xi,\xi \rangle \geq c_n\|\xi \|$. \eproof
\\

The sequence $\{b_n \}$ is needed in the case when the operator
$$ D_e=\sum_i E_id_{\hat{E}_i},$$
where $\hat{E}_i$ is an orthonormal frame of left translated vectorfields on one copy of $G$, has a non trivial kernel in the sense that the kernel is given by the span of $\{E_i \}_i$. This is clearly not the case for $U(1)$. In the appendix we will show that this is also not the case for $SU(2)$.

\begin{prop} \label{gaugeinv}
 When $D_b$ is constructed from a gauge admissible system of connections satisfying the demands in proposition \ref{selv}, $D_b$ is gauge invariant, i.e. $D_b=UD_bU^*$ for all $U\in \overline{\cg}^\triangle$. ($U$ acts trivially on the $Cl(1)$-part.)
\end{prop}

\textit{Proof.} From the proof of proposition \ref{megetgrim} it follows that $D_n^-$ is gauge invariant, i.e. invariant under $U_n$ for all $U_n\in \cg_n$. In particular the kernel of $D_n^-$ is  gauge invariant. Also the space (\ref{linie3}) is invariant under $\cg_n$ and therefore the projection $P_n$ onto the orthogonal complement of (\ref{linie3}) in the kernel of $D_n^-$ is gauge invariant under $\cg_n$. From this the invariance follows. \eproof \\

Note that the system of trivial connections with respect to the trivializations given by $\{\ce_i^{j,k}\}$ fulfills the the demands of proposition \ref{gaugeinv}.

\section{Appendix}
In this appendix we will first demonstrate the case of $U(1)$ to show what kind of growth conditions are needed on $a^{j,k}$. Secondly we show that for $SU(2)$ the perturbation with the $P_n$'s is not needed to obtain a semifinite spectral triple. 
\subsection{The $U(1)$-case}
We write $U(1)=\{e^{2\pi i\theta}|\theta \in [0,1]\}$ and choose the metric such that 
$$\langle \frac{d}{d \theta}, \frac{d}{d \theta}\rangle=1.$$
The system of connections we will use is the system of trivial connections. The operator $D_n$ has the form
$$\sum_{j\leq n,k,i}a^{j,k}\ce_i^{j,k} d_{\hat{\ce}_i^{j,k}}$$
Since $i$ can be only one in this formula, we will simply omit it. Also we will assume that $a^{j,k_1}=a^{j,k_2}$ for all $k_1,k_2$, and simply denote $a_j:=a^{j,k}$. All $\hat{\ce}^{j,k}$ commute. Hence
$$D_n^2=-\sum_{j\leq n,k}a_j^2(d_{\hat{\ce}^{j,k}})^2.$$
Since $D_n^2$  acts trivially in the Clifford bundle, and since the identity on the Clifford bundle is normalized to have trace $1$, we will omit the Clifford bundle in the rest of this computation.  

We will use the coordinate change $\Theta_n$ to rewrite $D_n^2$. The rewritten operator will be denoted $\tilde{D}_n^2$. According to lemma \ref{grimt} the vector fields
$$\hat{\ce}^{j,k}=c_1^{j,k}\partial_{\theta_1}+c_1^{j,k}\partial_{\theta_2} +c_2^{j,k}\partial_{\theta_3}+c_2^{j,k}\partial_{\theta_4}+\ldots +c_{2^{n-1}}^{j,k}\partial_{\theta_{2^n-1}}+c_{2^{n-1}}^{j,k}\partial_{\theta_{2^n}}, \quad j < n,$$
gets mapped to
$$ 2\sum_{l=1}c_l^{j,k}\partial_{\theta_1}+(c_1+2\sum_{l=2}c_l^{j,k})\partial_{\theta_2} + 2\sum_{l=2}c_l^{j,k}\partial_{\theta_3}+ \ldots + 2c_{2^{n-1}}^{j,k}\partial_{\theta_{2^n-1}}+ c_{2^{n-1}}^{j,k}\partial_{\theta_{2^n}} $$
under $\Theta_n$, and $\hat{\ce}^{n,k}$ gets mapped to  
$-\frac{1}{2 } \partial_{\theta_{2k}}$.
It follows from the construction that 
$$\sum_{j<n,k}|c_{2^{n-1}}^{j,k}|^2<1.$$
In particular for a function 
$$\xi (\theta_1,\ldots ,\theta_{2^n})=e^{2\pi i \theta_{2^n}l}, \quad l\in \bbZ$$
we see that 
$$\tilde{D}_n^2(\xi)=4 \pi^2l^2( \sum_{j< n,k}(c_{2^{n-1}}^{j,k} a_j)^2 +\frac{a_n^2}{4}) \xi,$$
and 
$$ \sum_{j < n,k}(c_{2^{n-1}}^{j,k}a_j)^2 +\frac{a_n^2}{4}\leq \max \{a_j^2\}_{j<n}+\frac{a_n^2}{4}.$$
From this follows, that $D$ only has finite many eigenvalues with finite multiplicity module  the semifinite trace in a bounded set of $\bbR$ if 
\begin{equation} \label{uet}
 a_n\to \infty.
\end{equation}
This is therefore a necessary condition for $D$ to have $\tau$-compact resolvent.

On the other hand the eigenfunctions for $\tilde{D}_n^2$ are on the form 
$$\exp (2\pi i (\sum_{l\leq 2^n} n_l \theta_l)).$$
The eigenfunctions for  $\tilde{D}_{n-1}^2$ are the functions on the form  
$$\exp (2\pi i (\sum_{l\leq 2^{n-1}} n_{2l-1} \theta_l)).$$
Hence the smallest new eigenvalue appearing from going to from $n-1$ to $n$ is bigger than $a_n^2$. Hence $D$ has $\tau$-compact resolvent if and only of condition (\ref{uet}) is satisfied. 

\subsection{The $SU(2)$-case}
We will show that the  Dirac operator on $SU(2)$ on the form 
$$ D_e=\sum_i E_id_{\hat{E}_i},$$
where $d$ is understood with respect to the trivialization given by $\{E_i\}$, has a trivial kernel in the sense that the kernel is given by the span of $\{E_i \}_i$. In other words the kernel consists of the constants with respect to the trivialization of the Clifford bundle given by  $\{E_i \}_i$. Accoording to the remark after the proof of \ref{hoved} this ensures that the perturbation with the $P_n$'s is not needed.

For the analysis of we will use that states $|jm\rangle $, where $j\in \{0,\frac{1}{2},1,\ldots \}$ and $m\in \{-j,-j+1,\ldots j-1,j\}$ form an orthonormal basis for $L^2(SU(2))$. 
Choose a basis $\hat{e}_1,\hat{e}_2,\hat{e}_3$ for $\mathfrak{su}(2)$ such that
$$[\hat{e}_i, \hat{e}_j]=\sum_k\varepsilon_{ijk}\hat{e}_k.$$
Define a metric on $\mathfrak{su}(2)$ by letting $\hat{e}_1,\hat{e}_2,\hat{e}_3$ be an orthonormal basis. Note this metric by left and right translation define a left and right invariant metric on $SU(2)$.

 The action of the corresponding left translated vectorfields of $\hat{e}_1,\hat{e}_2,\hat{e}_3$ is  best described by forming the raising and lowering operators
\[
\hat{e}_\pm = \hat{e}_1 \pm \rm{i}\hat{e}_2\;.
\]
The action of the translated vectorfields is given by
$$d_{\hat{E}_3}|jm\rangle =-im|jm\rangle,\quad d_{\hat{E}_\pm} |jm\rangle=-ic_\pm (m)|jm\pm 1\rangle,$$
where
$$c_\pm (m)= \sqrt{(j\mp m)(j\pm m+1)}.$$
We will restrict the operator $D_e$ to acting in one of the two irreducible representations of $Cl(T^*_{id}SU(2))$ instead of the full Clifford bundle. We chose the representation in $M_2$ given by
\[
E_1 = \left(
\begin{array}{cc}
0  & 1 \\
-1 & 0
\end{array}
\right)
\;,\quad
E_2 = \left(
\begin{array}{cc}
0  & i \\
i & 0
\end{array}
\right)
\;,\quad
E_3 = \left(
\begin{array}{cc}
i  & 0 \\
0 & -i
\end{array}
\right)
\;,
\]
The Dirac operator therefore has the form
\[
D_e=
\left( 
\begin{array}{cc}
0  & 1 \\
-1 & 0
\end{array}
\right)
\cdot d_{\hat{E}_1}
+
\left( 
\begin{array}{cc}
0  & i \\
i & 0
\end{array}
\right)
\cdot d_{\hat{E}_2}
+
\left( 
\begin{array}{cc}
i  & 0 \\
0 & -i
\end{array}
\right)
\cdot d_{\hat{E}_3}
\]
which is rewritten to
\[
D_e=
\left( 
\begin{array}{cc}
0  & 1 \\
0 & 0
\end{array}
\right)
\cdot d_{\hat{E}_+}
+
\left( 
\begin{array}{cc}
0  & 0 \\
-1 & 0
\end{array}
\right)
\cdot d_{\hat{E}_-}
+
\left( 
\begin{array}{cc}
i  & 0 \\
0 & -i
\end{array}
\right)
\cdot d_{\hat{E}_3}
\]
The square of $D_e$ is calculated
\ba 
D^2_e &=& - \left( 
\begin{array}{cc}
1  & 0 \\
0 & 0
\end{array}
\right)
\cdot d_{\hat{E}_+}d_{\hat{E}_-}
-
\left( 
\begin{array}{cc}
0  & 0 \\
0 & 1
\end{array}
\right)
\cdot d_{\hat{E}_-}d_{\hat{E}_+}
-
\left( 
\begin{array}{cc}
1  & 0 \\
0 & 1
\end{array}
\right)
\cdot d_{\hat{E}_3}d_{\hat{E}_3}
\nn\\&&
+
\left( 
\begin{array}{cc}
0  & i \\
0 & 0
\end{array}
\right)
\cdot d_{[\hat{E}_3, \hat{E}_+]}
+
\left( 
\begin{array}{cc}
0  & 0 \\
i & 0
\end{array}
\right)
\cdot d_{[\hat{E}_3, \hat{E}_-]}
\nn
\ea
Using how $\hat{E}_\pm$ and $\hat{E}_3$ act we get
\ba 
D^2_e
\left(\begin{array}{c}
a|jm\rangle\\
b|jm\rangle
\end{array}\right) &=& 
\left( 
\begin{array}{c}
a   \\
0 
\end{array}
\right)
\cdot (c_-)^2(m)|jm\rangle
+
\left( 
\begin{array}{c}
 0 \\
 b
\end{array}
\right)
\cdot (c_+)^2(m)|jm\rangle
\nn\\&&
+
\left( 
\begin{array}{c}
 -ib \\
 0
\end{array}
\right)
\cdot c_+(m)|jm+1\rangle
+
\left( 
\begin{array}{c}
0   \\
ia 
\end{array}
\right)
\cdot c_-(m)|jm-1\rangle
\nn\\&&
+
\left( 
\begin{array}{c}
a  \\
b
\end{array}
\right)
\cdot m^2|jm\rangle
\nn\\
&=&
\left( 
\begin{array}{c}
a   \\
b 
\end{array}
\right)
\cdot j(j+1)|jm\rangle
+
\left( 
\begin{array}{c}
a   \\
-b
\end{array}
\right)
\cdot m|jm\rangle
\nn\\&&
+
\left( 
\begin{array}{c}
-ib \\
0 
\end{array}
\right)
\cdot c_+(m)|jm+1\rangle
+
\left( 
\begin{array}{c}
0   \\
ia 
\end{array}
\right)
\cdot c_-(m)|jm-1\rangle
\nn
\ea
 For a given $j$ the following subspaces   
$$ \left( 
\begin{array}{c}
0   \\
b 
\end{array}
\right)
\cdot |j j\rangle
\quad 
\left(
\begin{array}{c}
a   \\
0 
\end{array}
\right)
\cdot |j -j\rangle
\quad 
\left(
\begin{array}{c}
 a  \\
0
\end{array}
\right)
\cdot |j m\rangle
+
\left(
\begin{array}{c}
0   \\
b 
\end{array}
\right)
\cdot |j m-1\rangle
$$
are invariant under $D_e^2$. Therefore analyzing $D^2_e$ is the same as analyzing matrices of the form
$$\left( 
\begin{array}{cc}
j(j+1)+m  & -ic_-(m) \\
ic_-(m) & j(j+1)-m+1
\end{array}
\right).$$
A lower bound for the product of the diagonal term is $j^4$. The off diagonal can be estimated by
$$c_-(m)^2=(j+m)(j-m+1)=j^2+j-m^2+m\leq j^2+2j. $$
Hence for $j\geq \frac{3}{2}$ these matrices are clearly invertible. For $j=1$ and $j=\frac{1}{2}$ it can easily be checked that these matrices are invertible. 
In particular we see that $D^2_e$ has trivial kernel.

Choosing the other irreducible representation of $Cl(T^*_{id}SU(2))$ corresponds to replacing $E_3$ with $-E_3$ and therefore $D^2_e$ in this representation also  has trivial kernel.

\bibliographystyle{plain}
\bibliography{ref}

\end{document}